\documentclass{article}

\usepackage[T1]{fontenc}
\usepackage[english]{babel}
\usepackage{lmodern}
\usepackage{graphicx}
\usepackage{amsmath}
\usepackage{url,hyperref,microtype}
\usepackage[toc,page]{appendix}
\usepackage{siunitx}
\usepackage{algorithm}
\usepackage{algorithmic}
\usepackage{authblk}

\title{Frequency-resolved dynamic functional connectivity and scale-invariant connectivity-state behavior}
\author[1,2]{Markus Goldhacker\thanks{markus.goldhacker@ur.de}}
\author[3,1]{Ana Maria Tom\'e}
\author[2]{Mark W. Greenlee}
\author[1,3]{Elmar W. Lang}
\affil[1]{CIML Lab, Dept. Biophysics, University of Regensburg, D-93040 Regensburg, Germany}
\affil[2]{Dept. Experimental Psychology, University of Regensburg, D-93053 Regensburg, Germany}
\affil[3]{IEETA, DETI, Universidade de Aveiro, P-3810-193 Aveiro, Portugal}
\date{}

\begin{document}

\maketitle

\begin{abstract}
Investigating temporal variability of functional connectivity is an emerging field in connectomics. Entering dynamic functional connectivity by applying sliding window techniques on resting-state fMRI (rs-fMRI) time courses emerged from this topic. We introduce frequency-resolved dynamic functional connectivity (frdFC) by means of multivariate empirical mode decomposition (MEMD) followed up by filter-bank investigations. We develop our method on the most canonical form by applying a sliding window approach to the intrinsic mode functions (IMFs) resulting from MEMD. We explore two modifications: uniform-amplitude frequency scales by normalizing the IMFs by their instantaneous amplitude and cumulative scales. By exploiting the well established concept of scale-invariance in resting-state parameters, we compare our frdFC approaches. In general, we find that MEMD is capable of generating time courses to perform frdFC and we discover that the structure of connectivity-states is robust over frequency scales and even becomes more evident with decreasing frequency. This scale-stability varies with the number of extracted clusters when applying $k$-means. We find a scale-stability drop-off from $k = 4$ to $k = 5$ extracted connectivity-states, which is corroborated by null-models, simulations, theoretical considerations, filter-banks, and scale-adjusted windows. Our filter-bank studies show that filter design is more delicate in the rs-fMRI than in the simulated case. Besides offering a baseline for further frdFC research, we suggest and demonstrate the use of scale-stability as a quality criterion for connectivity-state and model selection. We present first evidence showing that scale-invariance plays an important role in connectivity-state considerations. A data repository of our frequency-resolved time-series is provided.
\end{abstract}

\section{Introduction}
Functional connectivity is a key aspect in the analysis of rs-fMRI. It is based on calculating association measures -- mostly Pearson correlation -- between distinct regions in the brain. First attempts focused on the static case, for which whole time courses of resting-state-related brain regions were used for evaluating correlation coefficients representing the strength of their functional connections \cite{Eguiluz2005, Lang2012}. This approach resulted in many insights ranging from a small-world organization of brain graphs that are constructed from this so-called \textit{connectome} \cite{Bullmore2009} over deviations in functional connectivity between pathological and healthy brains \cite{Stam2007a,Ma2014b} to developmental changes of functional connectivity \cite{Geerligs2014}. It was also possible to identify similarities between physical systems like the \textit{Ising-model} of a ferromagnet and functional connectivity brain networks \cite{Fraiman2009}. However, using whole time courses integrates out all temporal dependences within the connectome, resulting in static average connectivities. Recently a paradigm shift occurred towards a so-called dynamic functional connectivity (dFC), which takes into account the temporal variability of functional connections in the brain. Investigating temporal fluctuations of functional connectivity thus has received considerable attention in the last few years \cite{Allen2012,Chang2010b,Cribben2012b,Calhoun2014,Deco2011b,Kopell2014}.

\cite{Allen2012} introduced a sliding-window technique applied to the time courses of independent components (ICs) resulting from a group independent component analysis (gICA) \cite{Calhoun2001} on a very large set of subjects undergoing rs-fMRI. This technique is also employed in the present study. The idea is to track the variability of correlation matrices formed from segments of the time courses of all ICs. Shifting the window one time step further, results in a new correlation matrix for the next time step with slight changes in correlation coefficients and so on. The vast set of correlation matrices resulting from such a sliding-window approach can be condensed into several representative correlation patterns by applying $k$-means clustering \cite{Allen2012}. These stable patterns, which robustly showed up as cluster-centroids, can be considered \textit{connectivity-states} given the fact that these centroids represent very robust and almost discrete correlation patterns reflecting characteristic connectivities that the brain goes through over time, while simultaneously remaining similar between subjects.

The number of extracted clusters has to be predefined using the $k$-means algorithm. Usually it is deduced from an elbow-criterion which indicates a sudden change in a cluster similarity index that, e.\,g., compares the variance within the extracted clusters to the variance between them. Recently, \cite{Leonardi2014b} applied a split-half reproducibility criterion to infer the proper number of connectivity-states in their study. Also scale invariance or scale-stability has been shown to be an inherent feature of rs-fMRI data \cite{Eguiluz2005,Kitzbichler2009,Moretti2013}. Scale-stability of connectivity-states should be taken into account when conducting a dFC analysis and we suggest to optimize it in terms of connectivity-state extraction. Consequently, in this study we explore the persistence of connectivity-states across frequency scales. 

One recent study \cite{Yaesoubi2015} investigated frdFC by applying a wavelet decomposition to the time courses resulting from gICA. Compared to this study, we do not rely on predefined temporal variations via chosen wavelets and we do not have to leave the classical dFC approach by delving into data-intrinsic frdFC by applying the sliding-window technique on data-inherent frequency scales. Nevertheless, resolving time-scales before entering the dFC pipeline seems promising and represents a new approach to dFC analysis. We address this point by applying multivariate empirical mode decomposition (MEMD), which is a data-driven method for extracting so-called IMFs \cite{Huang1998} which reveal inherent characteristic time scales of temporal variations of the quantities under study. Thus MEMD yields time courses corresponding to clearly separated and narrow-band frequency scales, which can be investigated individually by means of dFC. The frequency resolution and narrow-band frequency content of studied time courses, resulting from MEMD, are referred to as frequency scales in MEMD literature (see e.\,g. \cite{Looney2015,Ahmed2012,Costa2005}). Therefore we adopt this common term also for our study. Frequency scales can be defined in several ways in this context. Hence, we explore three different alternatives: (i) the canonical frequency scales inherent to IMFs resulting from MEMD; (ii) the uniform-amplitude frequency scales reached by removing amplitude information; (iii) cumulative scales generated by using MEMD as a coarse graining method. Since we are dealing with and exploiting the self-similarity of rs-fMRI time courses, starting with data-driven exploratory methods like MEMD is a natural way. \cite{Flandrin2004} propose to see EMD analyses as the means of choice when analyzing self-similar data. Our \textit{post hoc} analyses with filter-banks confirm that this way of approaching this kind of data is promising. Our aim is to offer a baseline for further frdFC research. Besides investigating the variable scale-stability of connectivity-states we find that the structure of connectivity-states is preserved over scales -- even for the lowest frequencies -- and we demonstrate the possible use of scale-stability as a quality criterion for connectivity-state definition and model selection.\footnote{Our results were presented in part as a talk and a poster at the annual meeting of the organization of Human Brain Mapping in 2015 (DOI: 10.13140/RG.2.1.2561.3929); abstract submission deadline was 01/15/2015. Thus we claim that we designed our study independently from upcoming publications on similar topics.}

\section{Material \& Methods}
\subsection{Data-set}
We based our analysis on volumetric data from the preselected bundle of 100 unrelated human subjects from the \ref{fig:scaleStability_robustnessK5}00 release of the \textit{Human Connectome Project} \cite{VanEssen2012}, in which each subject went through four rs-fMRI sessions lasting 14min 33s resulting in 1200 volumes per session and $n = 400$ sessions in total. Data was acquired at customized 3T MRI scanners at Washington University using multi-band (factor 8) acquisition techniques \cite{Moeller2010, Feinberg2010, Setsompop2012, Xu2012}. From the different versions of the data we chose the most preprocessed data set with motion-correction, structural preprocessing, and ICA-FIX denoising \cite{Glasser2013, Jenkinson2002, Jenkinson2012, Fischl2012, Smith2013, Salimi-Khorshidi2014, Griffanti2014}. The rs-fMRI data has a TR of $720$ms and TE of $33.1$ms acquired by a Gradient-echo EPI sequence. Flip angle was 52$^{\circ}$ and the FOV 208 mm $\times$ 180 mm with a slice thickness of 2mm, 72 slices, and isotropic 2mm voxel size. As additional preprocessing we applied a Gaussian smoothing kernel with a FWHM of 5mm using SPM8 software package\footnote{http://www.fil.ion.ucl.ac.uk/spm/} and we discarded the first five scans of each session from our analysis.

\subsection{Group-ICA}
To apply gICA \cite{Calhoun2001} on our data we used the GIFT toolbox\footnote{http://mialab.mrn.org/software/gift/}. On the single-subject level, our data matrices $\mathbf{D}_{i,M\times K}$ have time as the row dimension ($M = 1195$) and voxel-space as the column dimension ($K = 193965$). As additional preprocessing, before entering the analysis pipeline, we used variance normalization by linearly detrending and $z$-scoring each voxel time-series. At the single-subject reduction step, we extracted $m = 45$ principal components (PCs) from the data of each session. By solving the eigenvector decomposition $\mathbf{C}_i = \mathbf{V}_i\mathbf{L}_i\mathbf{V}_i^T$ of the data-covariance matrix $\mathbf{C}_i = \mathbf{D}_i^T\mathbf{D}_i$ with $\mathbf{V}_i$ representing the matrix of PCs and $\mathbf{L}_i$ the eigenvalue matrix, a PCA allowed to substantially reduce the dimensionality of the problem. By using the $m = 45$ projections onto the PCs with the highest eigenvalues from $\mathbf{L}_i$, a $m \times K$ reduced data matrix $\mathbf{R}_i^{(s)}$ resulted on the single-subject level. This was followed by a group reduction step, stacking all $\mathbf{R}_i^{(s)}$ in the row domain resulting, on the group level, in an $nm\times K$ matrix $\mathbf{R^{(g)}}$ entering once again a PCA, in analogy to the single-subject step, and, finally, extracting 30 PCs across all subjects. All PCAs were, because of limited on-board memory, implemented with the help of an expectation maximization algorithm \cite{Roweis1998}. On the resulting group data set, we evaluated gICA $\mathbf{R}^{(g)} = \mathbf{T}\mathbf{S}$ by applying the Infomax algorithm \cite{Bell1995} to extract spatially independen components (ICs) $\mathbf{S}$ and by applying the ICASSO approach \cite{Himberg2004} for component stability while repeating the ICA algorithm 10 times. Finally, ICs were extracted from our data similar to \cite{Smith2013}, which ensures computability when it comes to applying the MEMD algorithm \cite{Huang1998,Mandic2013}. To come back from the group level to the single-subject level, the  back-reconstruction algorithm GICA3 \cite{Erhardt2011a} was employed yielding subject specific time courses $\mathbf{T}_i(t)$ and spatial maps $\mathbf{S}_i$. Afterwards, the spatial maps were $z$-scored. The results of the gICA analysis are illustrated in figure \ref{fig:gicaResEMD_appl}A.

We tested our results extensively concerning stability and performed our analysis with different numbers of components. Looking at the dynamic range and low to high frequency ratio \cite{Allen2011}, which reflects spectral characteristics of the ICs, we find that the obvious artifact ICs 7, 15, and 17 have lower values in both measures (fig. \ref{fig:spectrumCharacteristics}). Additionally ICs 10, 12, and 13 have lower values in those measures than the best artifact component (IC 7). Consequently, we tested our approach by dividing the set of ICs in a more conservative, an intermediate, and a liberal set of resting-state networks (RSNs), by which we identify ICs representing functionally important networks. For the conservative set we removed ICs 10, 12, and 13, in addition to the three artifactual ICs related to vascular (IC 15), cerebro-spinal fluid (IC 7), and white matter components (IC 17) resulting in 24 ICs as RSNs. The more liberally selected set of ICs consisted of a total of 27 RSNs leaving out the artifact ICs, while the intermediate selection led us to remove IC 13 in addition to the artifactual ICs, since dFC analysis showed that this IC is very noisy resulting in a data set with 26 ICs. We evaluated our method for each data set type and find that results look very similar over all types with the clearest results for the conservative run. Following the classification scheme used in the study of \cite{Allen2011} it is plausible that the conservative data set gives the best and most valid results, since the ICs added in the intermediate and the liberal data set have worse spectral characteristics than the best artifact IC. As our method turned out to be stable concerning the three types of datasets, we show results of the conservative data set, since we want to deduce our results from data with the least amount of noise and RSNs having time courses with spectral characteristics at least better than artifactual ICs.

\subsection{Multivariate empirical mode decomposition}

After having decomposed our data set into a set of independent components consistent across a group of subjects, the next step then applied, for each session $i$ separately, an MEMD on these time courses $\mathbf{T}_i(t)$. MEMD represents a data-driven approach of decomposing e.\,g. non-stationary and non-linear time-series thus serving as a suitable tool for the analysis of brain data time courses. Following we present the concept of MEMD based on the canonical empirical mode decomposition (EMD) method. The approach was first introduced by \cite{Huang1998} and later extended to a noise-assisted ensemble EMD (EEMD) by \cite{Wu2009b}. The decomposition results in  IMFs, which represent characteristic inherent modes of the univariate time course under consideration. Let $x(t)$ be a general univariate time-series, then EMD extracts one dimensional inherent modes $u^f(t)$ such that the original signal can be expanded into these modes plus a residual non-oscillating trend $r(t)$ \cite{Huang1998,Mandic2013}
\begin{equation}
x(t) = \sum_{f = 1\\}^{F} u^f(t) + r(t)
\label{eq:pEMD}
\end{equation}
Note that EMD results in a complete decomposition of the signal, i.\,e., summing up all IMFs and the residue results in the original time course. Thus in contrast to exploratory signal decomposition techniques referred to above, absolute values of component amplitudes are of relevance giving each time point a unique partner over all IMF-indices. EMD starts by selecting all maxima and minima of $x(t)$ and creates an envelope by spline interpolation for the sets of maxima and minima separately. Afterwards the mean of the envelopes is subtracted from $x(t)$ and it is checked, if the resulting time course meets the criteria for being an IMF -- this process is called \textit{sifting}. The two criteria for a time course for being an IMF are: having symmetrical upper and lower envelopes and the number of extrema and zero-crossing differing at most by one \cite{Mandic2013,Wang2010}. If it does not meet the criteria, the process is started again with this new time course. Having extracted one IMF leads to the subtraction of the latter from the original time course, and the sifting process starts again with the remaining time course until a non-oscillating function is left, which is considered the residue $r(t)$. By means of the so-called Hilbert-Huang-transformation the instantaneous frequency and amplitude $a^f(t)$ can be extracted for each IMF. \cite{Looney2015} suggests to also explore uniform-amplitude IMFs in terms of frequency scales
\begin{equation}
\hat{u}^f(t_i) = \frac{u^f(t_i)}{a^f(t_i)},
\label{eq:uniformAmp}
\end{equation}
where for each time point $t_i$ the components of the multivariate IMF $u^f(t_i)$ are divided by its corresponding amplitude $a^f(t_i)$ resulting in uniform-amplitude IMFs $\hat{\mathbf{U}}^f(t)$. They found in an MEMD decomposition of EEG time-series resulting from a steady-state visual evoked potential task that removing amplitude information from IMFs reveals the true regularity of the signal. Therefore, we explore uniform-amplitude IMFs as alternative frequency scales in terms of frdFC.

\cite{Mandic2013} mention at least two popular shortcomings of plain EMD. First, it is not ensured that modes appear in just one IMF, rather they could spread over several IMFs. This condition is known as \textit{mode mixing}. Second, they mention a problem with so-called \textit{end effect artifacts}, indicating that creating proper envelopes needs a sufficient number of extrema. As time courses, being decomposed, have finite range, the density of extrema tends to decrease near the edges of the sampled time interval. Therefore the fit of the envelopes is more error-prone at the beginning and end of the time courses.

\begin{figure}[h!]
\centering
\includegraphics[width = 1\textwidth]{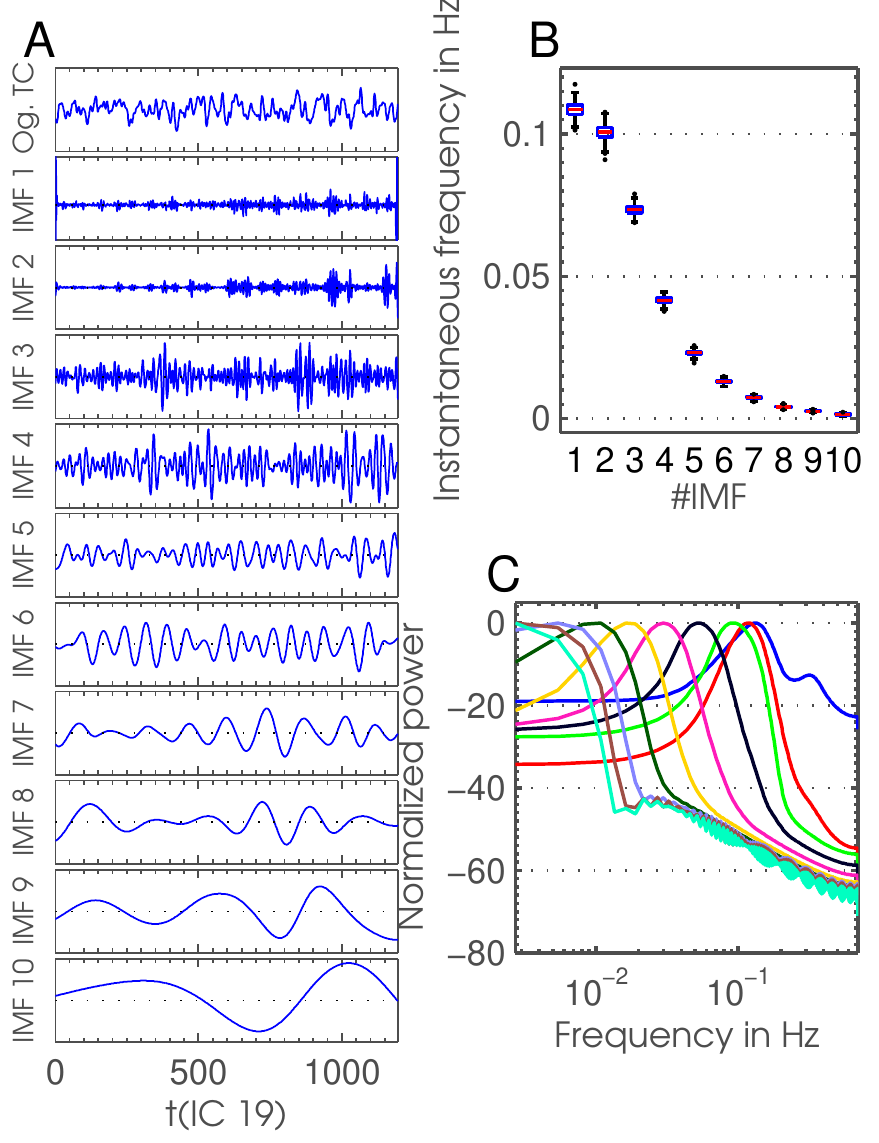}
\caption{(A) The original time course and time courses of IMFs $1-10$ of the first IC of one session (session 322) are shown as an example. (B) This panel shows box-plots for the instantaneous frequency against IMF indices. Data points for calculating the box-plots represent the instantaneous frequency averaged over time points and ICs. Black dots represent outliers concerning box-plots. MEMD mode alignment yields very small deviations in frequency between sessions, which enables us to do dFC on different frequency scales. (C) Depicted here are the power spectrum densities (normalized) for all IMF indices used. Each curve represents the average power spectrum densities over all ICs, sessions and subjects. The abscissa is in log-scale and shows the frequency range $0.0027\mbox{Hz} < f < 0.6944\mbox{Hz}$.}
\label{fig:exampleIMFs_PSD_boxPl}
\end{figure}

A further improvement of canonical EMD is provided by its extension towards a multivariate, noise-assisted ensemble EMD (MEMD) \cite{Mandic2013, Rehman2011b, Rehman2009}. In short, EMD is extended to interpret input of multiple channels as a multi-variate signal. Compared to the noise-assisted variant EEMD, noise-channels are added to the multi-dimensional signal in the MEMD case instead of adding noise onto the time courses themselves. Creating an ensemble of IMFs by re-doing MEMD with newly generated noise as several realizations yields the noise-assisted ensemble extension of MEMD. Strictly speaking, the IMFs $\mathbf{U}^f(t)$ extracted by MEMD are multivariate with the same dimensionality as the multivariate signals  themselves. After creating an ensemble, representatives of IMFs are created by averaging over all realizations thereby reducing the mentioned shortcomings common to plain EMD \cite{Rehman2013b}. One of the outstanding features of MEMD is its accurate mode-alignment quality, which renders it most suitable for a frdFC analysis. This is because for applying a sliding-window approach on separate frequency scales over many subjects and RSNs, the frequencies need to be aligned accurately. Therefore we use the MEMD approach for the decomposition of the time courses $\mathbf{T}_i(t)$ of our data set resulting from the back-reconstruction step. MEMD is applied to $\mathbf{T}_i(t)$ of each session. Figure \ref{fig:exampleIMFs_PSD_boxPl}A illustrates, as an example, time courses of the first IC from 10 IMFs of one session. The instantaneous frequencies of the modes decrease with increasing IMF-index. The box-plots of the frequency range of the IMF-indices (fig. \ref{fig:exampleIMFs_PSD_boxPl}B) and the corresponding spectral power densities (fig. \ref{fig:exampleIMFs_PSD_boxPl}C) show that mode mixing was avoided by separating frequencies of different scales. In particular, figure \ref{fig:exampleIMFs_PSD_boxPl}B shows the distinct quality of MEMD to align modes over sessions and RSNs.

Before performing frdFC introduced below and applying MEMD on the time courses of the RSN ICs $\mathbf{T}_i(t)$, a low-pass filter with a cut-off of 0.15Hz, despiking, and cubic detrending was applied on the time courses of the RSN ICs.  This means that the time courses had the same level of preprocessing both before entering dFC, and MEMD. On the preprocessed time courses, we applied MEMD by using scripts\footnote{http://www.commsp.ee.ic.ac.uk/$\sim$mandic/research/emd.htm} provided by \cite{Rehman2009} after adapting them to the noise-assisted ensemble approach by adding four white Gaussian noise channels with a power of $6\%$ \cite{Rehman2013b} of the average power of the original signal to the 24 time courses $\mathbf{T}_i(t)$ of the RSN ICs in 30 realizations of noise for all 400 sessions. These noise channels were then discarded from all IMFs. For our further analysis, we used the first 10 IMFs only and discarded IMFs with higher indices. In some sessions and certain realizations, the algorithm stopped after IMF-index 10, which means that IMF 10 represented the residual signal. To assure comparability, we repeatedly initialized the algorithm with newly generated noise channels until it extracted at least 11 IMFs (10 IMFs + 1 residue), since this number was most common across all sessions and realizations. With this procedure we ensured the same stopping criterion throughout all sessions and that no residue was taken into account in calculations with time courses of IMF index 10. This procedure results in frequency-resolved time courses, or rather multivariate IMFs $\mathbf{U}^f_i(t)$ for each session $i$ on frequency scales $f = 1,\dots, F$ with $F = 10$. $\mathbf{U}^f_i(t)$ are introduced as the canonical frequency scales resulting from MEMD further being investigated by means of frdFC.

The last frequency scale we introduce emerged from so-called intrinsic multivariate multiscale entropy research (iMMSE) \cite{Ahmed2012}, which is an extension of plain sample entropy \cite{Richman2000}, multiscale entropy \cite{Costa2005}, and multivariate multiscale entropy \cite{Ahmed2011} to the intrinsic form using MEMD decomposition. Sample entropy quantities measure the complexity of time courses. iMMSE operates in a data-driven manner on different scales of the multivariate time-series. To reach different frequency scales of the time courses of interest the IMFs resulting from MEMD are cumulated
\begin{equation}
\mathbf{U}_{cum}^f(t) = \sum_{j = f}^{F} \mathbf{U}^f(t).
\label{eq:cumIMFs}
\end{equation}
This implies that $\mathbf{U}_{cum}^1(t)$ is the original time-series and all following $\mathbf{U}_{cum}^f(t)$ are reduced versions of it consecutively taking IMFs out of it, which results in cumulative frequency scales. This resembles a coarse graining, but in a data-driven manner.

To sum up, we look at three different literature-based definitions of frequency scales resulting from MEMD to investigate frdFC: (i) the canonical scales represented by the original IMFs resulting from MEMD; (ii) the uniform-amplitude scales resulting from removing amplitude information from the IMFs; (iii) the cumulative scales common to iMMSE literature. Our method is developed on the canonical scales and then transfered to the other types of scales.

\subsection{Dynamic Functional Connectivity}
\begin{figure}[h!]
\centering
\includegraphics[width = 1\textwidth]{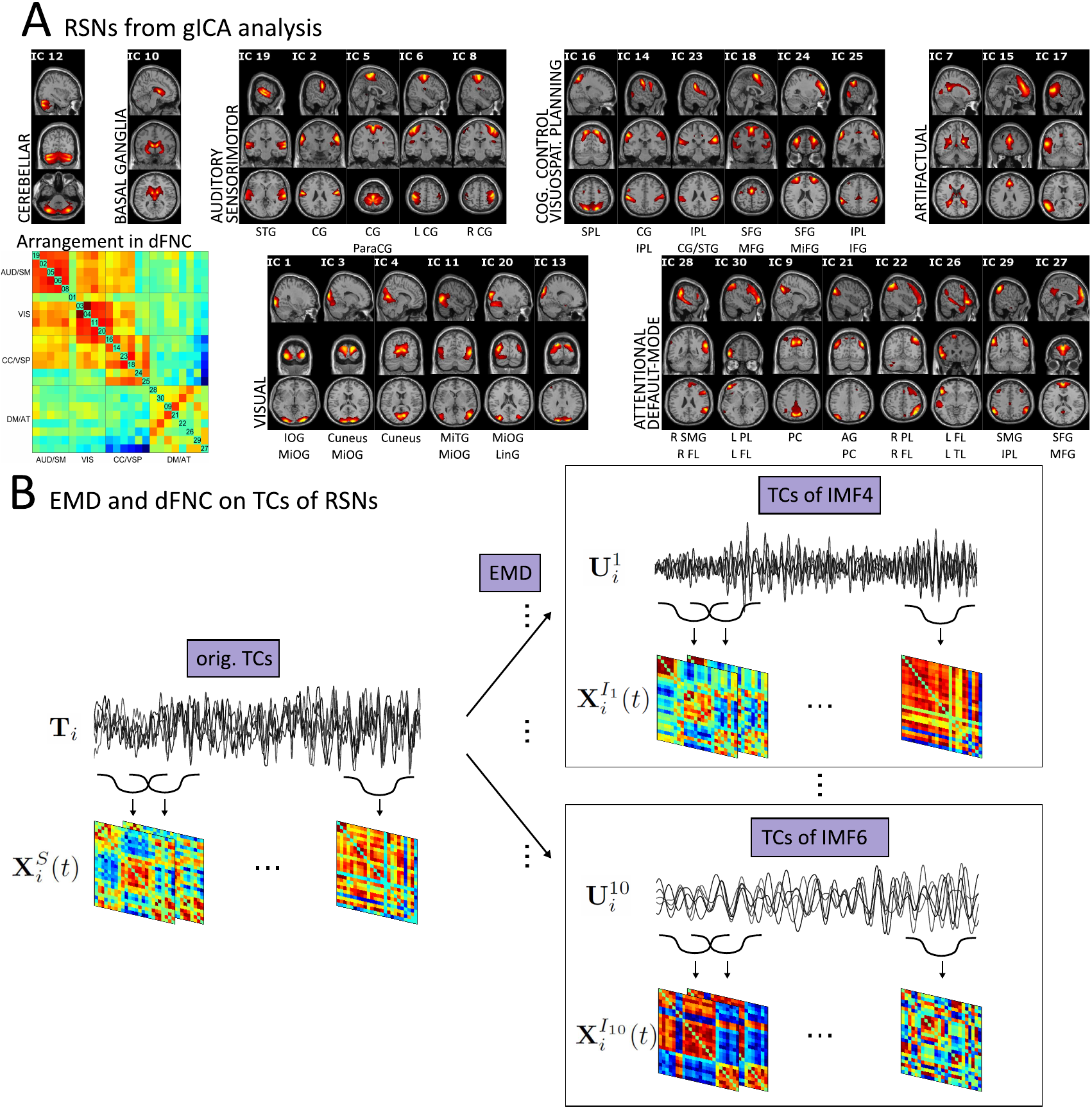}
\caption{(A) This panel shows the results from the gICA analysis of all 400 sessions arranged in modules concerning similar correlation behavior. Activity maps represent the average over all sessions with values being $z$-transformed. The cut-off is chosen as $z > 2$ and the highest value of the color-range is individually determined for each map. The arrangement of the RSNs in the dFC matrices is shown on bottom left of this panel. This correlation matrix shows the average of the dFC matrices over all time points and sessions. Note the clear segregation into functional modules. Below each RSN characteristic areas are mentioned. Abbreviations: Superior Temporal Gyrus (STG), Central Gyrus (CG), Superior Parietal Lobule (SPL), Inferior Parietal Lobule (IPL), Superior Frontal Gyrus (SFG), Medial Frontal Gyrus (MFG), Middle Frontal Gyrus (MiFG), Inferior Frontal Gyrus (IFG), Inferior Occipital Gyrus (IOG), Middle Occipital Gyrus (MiOG), Middle Temporal Gyrus (MiTG), Lingual Gyrus (LinG), Supramarginal Gyrus (SMG), Frontal Lonule (FL), Parietal Lobule (PL), Precuneus (PC), Angular Gyrus (AG), Temporal Lobule (TL), Superior Frontal Gyrus (SFG). (B) The base of our approach is to decompose the time courses of all RSNs by EMD resulting in ten separate frequency scales. To both on the original time courses and the time courses of each frequency scale dFC is applied yielding sets of correlation matrices. For reaching frequency scale analysis, sets of correlation matrices from the original time courses and the time courses resulting from the MEMD decomposition are needed (see text for details).}
\label{fig:gicaResEMD_appl}
\end{figure}

For our analysis we needed to evaluate the dFC approach on the single-subject time courses of our data set $\mathbf{T}_i(t)$ and on their corresponding frequency-resolved time courses $\mathbf{U}^f_i(t)$ (fig. \ref{fig:gicaResEMD_appl}B). We resolve signal time courses on various frequency scales to gain deeper insight into dFC networks and the dynamics of their related connectivity-states. For the whole study, the following parameters concerning dFC with constant window size hold. We use a window size for the sliding-window approach of $80$ TRs resulting in a window length of $\SI{57.6}{\s}$. For our approach we had to use boxcar windows instead of tapered windows (explanation see section \ref{sec:adjGIFT} in the supplementary material). The time steps between each window are chosen to be $1$ TR resulting in $1115$ time steps for each session. Correlation matrices for each window were calculated using L1 regularization with 10 repetitions estimating precision matrices \cite{Varoquaux2010, Smith2011} by applying graphical LASSO \cite{Friedman2008}. We ended up with sets of correlation matrices $\mathbf{X}_i^S(t)$ from the standard time courses $\mathbf{T}_i(t)$ and sets of correlation matrices $\mathbf{X}_i^{I_f}(t)$ from time courses $\mathbf{U}^f_i(t)$ of IMF index $f$ for each session. Every set of correlation matrices from one session had the dimension $C \times C \times T$, with $C = 24$ being the number of used RSNs and $T = 1115$ the number of windows. To stabilize variance, we applied a Fisher-transformation on every correlation coefficient before $k$-means clustering was performed. The $k$-means algorithm, implemented in the GIFT toolbox, was applied to the set of correlation matrices from the original time courses $\left\{\mathbf{X}_i^S(t)\right\}_{i = 1,\dots, n}$ and the sets of correlation matrices from the time courses resulting from the EMD $\left\{\mathbf{X}_i^{I_f}(t)\right\}_{i = 1,\dots, n}$ on each frequency scale separately with a maximum of 500 iterations and $k\in[2;10]$. These sets of correlation matrices that entered the clustering consisted of correlation matrices from all sessions. From each of the $10$ frequency scales, and from the original time courses, $400\times 1115$ correlation matrices resulted by applying dFC. On each of those sets of correlation matrices, $k$-means clustering was applied with $k$ ranging from 2 to 10. Afterwards, Fisher-transformed correlation coefficients were back-transformed to original values for our analyses.

We grouped our RSNs with respect to their similarity in the average correlation matrix (see arrangement in fig. \ref{fig:gicaResEMD_appl}A, left). This yielded an auditory/sensorimotor (AUD/SM), a visual (VIS), a cognitive control/visuospatial planning (CC/VSP), and a default mode/attentional module (DM/AT). So dividing our RSNs resulted in a clear modularization of the connectome with common functional properties of the members and distinct connectivity characteristics between different modules.

\section{Results}
\subsection{Robustness of connectivity-states over a wide frequency range}
As introduced above, the dFC-analysis applied to the inherent time courses offered by the MEMD results in corresponding sets of inherent correlation matrices. We investigate the dynamics of connectivity-states, identified as centroids of a $k$-means clustering procedure, like in a standard dFC analysis. But, due to an MEMD pre-processing, we can analyze the dynamics on different inherent time scales as represented by the local frequencies and as identified through the IMFs extracted from the signals. To investigate the robustness of the connectivity-states across different inherent frequency scales as identified by MEMD, we developed an ordering algorithm (see alg. \ref{alg:orderUnique} in the appendix) to line up, across different frequency scales, most similar connectivity-states.

\begin{figure}[h!tbp]
\centering
\includegraphics[width = 1\textwidth]{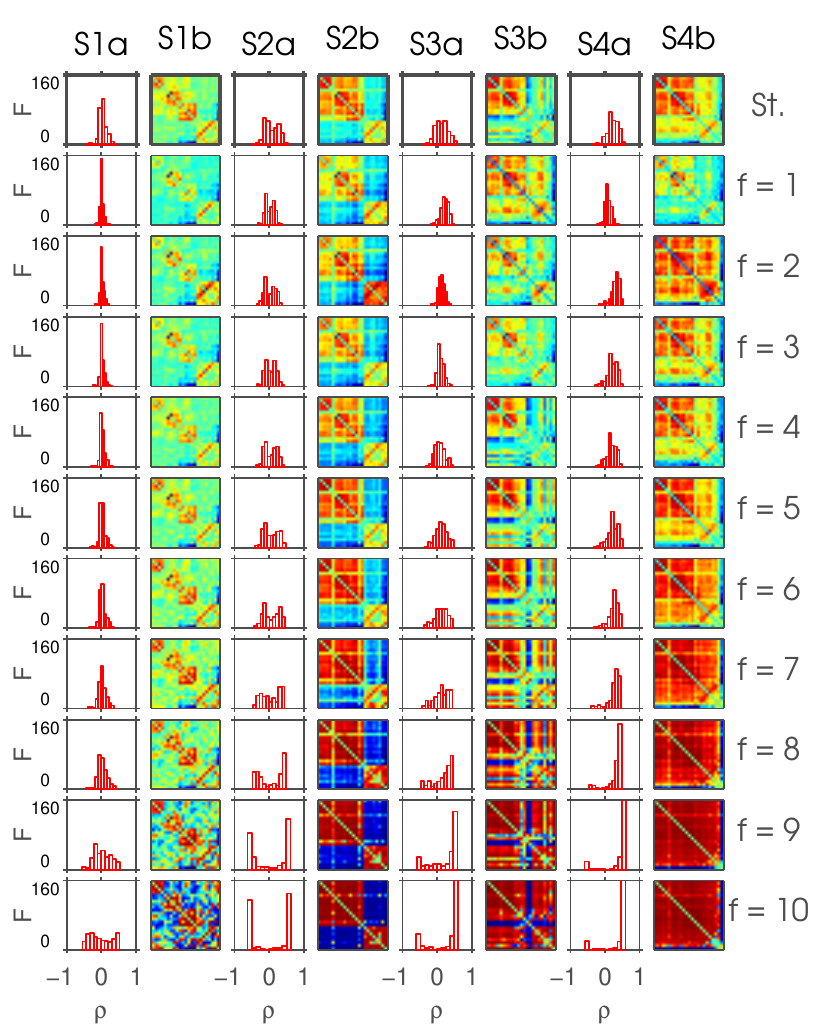}
\caption{Depicted is the result of the ordering algorithm \ref{alg:orderUnique} on the $k$-means run with $k = 4$. Connectivity-states are shown in columns with suffix *b and the coloring is individually adjusted by MATLAB for each plot to range from minimum (blue) to maximum (red) value to emphasize the structural similarity of connectivity-states over frequency scales. The absolute frequency $F$ of correlation coefficients $\rho$ are plotted for corresponding connectivity-states in columns with suffix *a. The first row shows the connectivity-states from the standard dFC approach arranged by their occurence frequency decreasing from left to right. From the second to the last row the connectivity-states for each frequency scale are shown. This is the result of the ordering algorithm \ref{alg:orderUnique} for increasing frequency scales.}
\label{fig:scaleStability_robustnessK4}
\end{figure}

Applying $k$-means clustering separately to each set of correlation matrices of intrinsic modes with their related inherent frequency scales, results in an array of plots of dimension $F\times k$ (see e.\,g. fig. \ref{fig:scaleStability_robustnessK4}) with $F$ extracted IMFs and $k$ states according to the number $k$ of underlying centroids. With varying $k\in [2;10]$, this procedure resulted in nine plot arrays. The algorithm sorts connectivity-states across all frequency scales according to closest similarity (see fig. \ref{fig:scaleStability_robustnessK4} and figs. \ref{fig:scaleStability_robustnessK2} -- \ref{fig:scaleStability_robustnessK10} in the supplementary material). The algorithm is initialized several times while shuffling frequency scales to avoid any sequence effects. However, for visualization purposes and a qualitative analysis, we illustrate the results of the realization with ordered frequency scales.

With $k = 2$, mainly results on lower frequency scales differ strongly from results on higher scales. Extracting  $k = 3$ and $k = 4$ connectivity-states, respectively, (see fig. \ref{fig:scaleStability_robustnessK3} and \ref{fig:scaleStability_robustnessK4}) reveals a very robust alignment across all frequency scales. Note that with $k = 4$ connectivity-states, another robustly aligned column, i.\,e. state, is added to the three connectivity-states already obtained from the run with $k = 3$. For larger number $k$ of clusters, this is no longer the case, since newly added columns become increasingly misaligned reflecting a larger diversity among connectivity-states.

Visual inspection of the aligned connectivity-states is already encouraging. Connectivity-states, resulting from the dFC standard approach (first row in all plots), are arranged by their frequency-of-occurrence. We report connectivity-states of figure \ref{fig:scaleStability_robustnessK4} in more detail, and identify them as being similar to states 1-4 from \cite{Allen2012} where the discussion of the physiology of those states can be found. The first connectivity-state (\ref{fig:spectrumCharacteristics}b) in all plots corresponds to their state 1 relating to static functional connectivity, which also holds for our results even over the whole frequency range in almost all $k$-means runs. The second state, which they refer to as state 2 (\ref{fig:corrMatIMFs}b in figs. \ref{fig:scaleStability_robustnessK4}, \ref{fig:scaleStability_robustnessK5}, \ref{fig:scaleStability_robustnessK6}; \ref{fig:scaleStability_robustnessK2}b in figs. \ref{fig:scaleStability_robustnessK3}, \ref{fig:scaleStability_robustnessK8}, \ref{fig:scaleStability_robustnessK9}, \ref{fig:scaleStability_robustnessK10}), is aligned robustly across all scales and groupings, except for $k = 7$ and $k = 2$. This state also has a very typical pattern often revealed by standard dFC analysis. The RSNs of the DM/AT network tend to anti-correlate with all other RSNs. The next connectivity-state (\ref{fig:corrMatIMFs}b in figs. \ref{fig:scaleStability_robustnessK3}; \ref{fig:scaleStability_robustnessK2}b in fig. \ref{fig:scaleStability_robustnessK6}; \ref{fig:scaleStability_robustnessK3}b in figs. \ref{fig:scaleStability_robustnessK4}, \ref{fig:scaleStability_robustnessK5}; \ref{fig:scaleStability_robustnessK7}b in figs. \ref{fig:scaleStability_robustnessK7}, \ref{fig:scaleStability_robustnessK10}; \ref{fig:scaleStability_robustnessK9}b in fig. \ref{fig:scaleStability_robustnessK9}) is similar to their state 3, also showing the largest slope in the linear fit of the frequency of occurrences over time. This aspect, and the lack of anti-correlation of the DM/AT RSNs, suggests that we are dealing with a state of drowsiness becoming more probable with increasing scanning time. The last state, which we want to mention, is connectivity-state \ref{fig:scaleStability_robustnessK2}b in figure \ref{fig:scaleStability_robustnessK4} relating to their state 4, It shows a  pattern distinct to the other three states and can also be found in other $k$-means runs (\ref{fig:scaleStability_robustnessK2}b in fig. \ref{fig:scaleStability_robustnessK5}; \ref{fig:scaleStability_robustnessK3}b in fig. \ref{fig:scaleStability_robustnessK6}, \ref{fig:scaleStability_robustnessK8}, \ref{fig:scaleStability_robustnessK9}; \ref{fig:scaleStability_robustnessK6}b in fig. \ref{fig:scaleStability_robustnessK10}). In this connectivity-state, RSNs $24$ and $25$ tend to anti-correlate with AUD/SM- and VIS-RSNs. Taking a closer look at the average correlation matrix in figure \ref{fig:gicaResEMD_appl}A and the classification of the modules reveals that this behavior results in a separation of the CC/VSP from the VIS module. RSNs $24$ and $25$ loose their connection to the VIS module and switch to anti-correlation. As with all other connectivity-states, this state's structure tends to become more evident with decreasing frequency, especially when a proper $k$ with high scale-stability is chosen.

\subsection{Variability of scale-stability of connectivity-states}
\cite{Kitzbichler2009} investigated broadband criticality of the resting-state and could identify scale invariance over a wide frequency range. The former is an indicator of self-organized criticality and links resting-state data to physical systems in a critical state. This result suggests that looking in rs-fMRI for scale invariance, i.~e. consistency of parameters over a wide frequency range, is promising.

\begin{figure}[h!tbp]
\centering
\includegraphics[width = .8\textwidth]{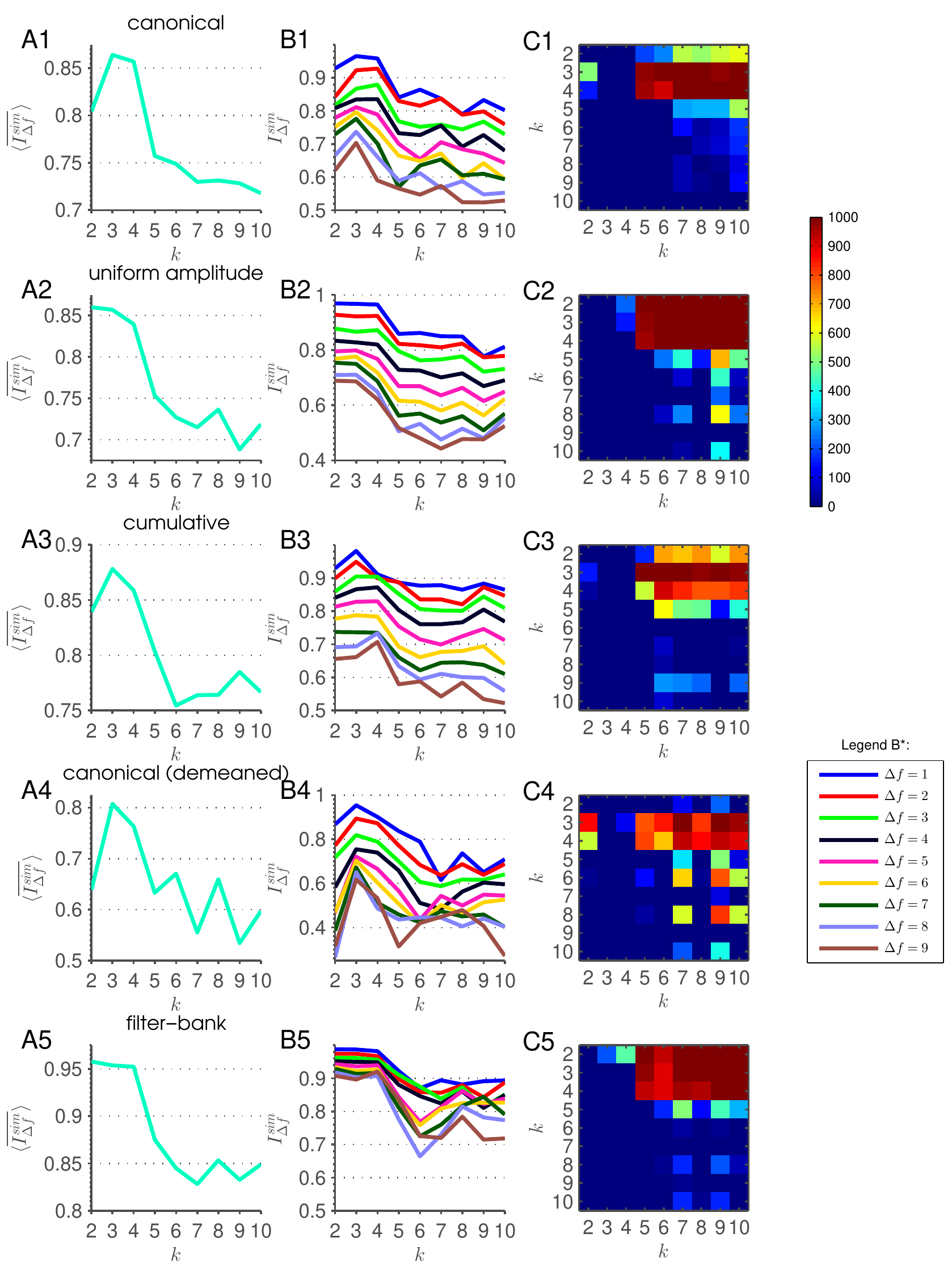}
\caption{This figure summarizes the results of the scale-stability analysis for the three frequency scale definitions under consideration and the result of the component-wise temporal demeaning for the canonical scales: (*1) canonical scales; (*2) uniform-amplitude scales; (*3) cumulative scales; (*4) canonical scales; (*5) filter-bank. (B*) Using spatial correlation of the aligned connectivity-states over increasing scale distances $\Delta f$ results in similarity measure $I_{\Delta f}^{sim}(k)$. (A*) Averaging over all frequency scales and all shuffled realizations in the ensemble (see text for details) results in a global similarity measure $\langle \overline{I_{\Delta f}^{sim}}\rangle(k)$, which can be plotted against $k$. The shuffling of frequency scales was introduced to avoid any sequence effects or weighting of certain frequency scales. The ordering algorithm was initiated 1000 times with shuffled scales for each iteration. (C*) \textit{Post hoc} two-sample signed Wilcoxon rank-sum tests (significance level: $.00196$, right-tailed) revealed that $k$-means with $k\in[3;4]$ result in significantly higher $\langle \overline{I_{\Delta f}^{sim}}\rangle(k)$ than $k$-means with $k\in[5;10]$ in most of the 1000 iterations with a separation between the two regimes for the canonical and uniform-amplitude approach. Depicted are the number of significant Wilcoxon rank-sum tests for each possible pairing. The results for cumulative scales are not as clear. Uniform-amplitude and filter-bank scales even show plateau behavior of $\langle \overline{I_{\Delta f}^{sim}}\rangle(k)$.}
\label{fig:scaleStabilityIndex}
\end{figure}

Looking at connectivity-states across frequency scales more quantitatively reveals that connectivity-states, deduced from particular $k$-means groupings, are highly stable concerning their similarity. The similarity of states can be looked at more closely by correlating  connectivity-states across adjacent frequency scales. After aligning connectivity-states for each $k$-means run, patterns at frequency scales $f$ and $f + 1$ are correlated -- thus $\Delta f = 1$ in the sense of frequency "distance". Then $\Delta f$ is increased step by step to look for similarity at more distant frequency scales. Finally, the global measure of similarity is plotted for each $\Delta f$ (see fig. \ref{fig:scaleStabilityIndex}B1).

Correlation is determined by looking for highest association between patterns. The latter is estimated by averaging all correlation coefficients resulting from the comparison of states across all frequency 'distances'. By doing so we achieve a measure of similarity for each $k$ and each $\Delta f$ according to

\begin{equation}
I_{\Delta f}^{sim}(k) = \frac{1}{(F - \Delta f)k}\sum_{i = 1}^k\sum_{f = 1}^{F  - \Delta f}\mbox{corr}\left(\mathbf{X}^f_i, \mathbf{X}^{f + \Delta f}_i\right).
\label{eq:simOverScales}
\end{equation}

Averaging $I_{\Delta f}^{sim}$ over $\Delta f$ results in a global similarity measure $\overline{I_{\Delta f}^{sim}}$ for a complete $k$-means run, i.\,e. for each $k$. This measure takes into account both similarity over nearby as well as distant frequency scales. To avoid sequence artifacts or any weighting of scales, the ordering algorithm shuffles frequency scales before ordering, and shuffling is repeated $1000$ times. Each time, the scale-stability index of equation \ref{eq:simOverScales} is calculated and afterwards averaged over the $1000$ iterations. Plotting this ensemble average $\langle \overline{I_{\Delta f}^{sim}}\rangle(k)$ for each $k$, results in a high similarity across frequency scales of connectivity-states for small $k\in[2;4]$, but similarity decreases for larger numbers of extracted states corresponding to  $k\in[5;10]$ (fig. \ref{fig:scaleStabilityIndex}A1). This decline of $\langle \overline{I_{\Delta f}^{sim}}\rangle(k)$ from $k = 4$ to $k = 5$ can be seen by looking at column \ref{fig:scaleStability_robustnessK5}b in figure \ref{fig:scaleStability_robustnessK5}. Extracting $k = 5$ connectivity-states results in four  clusters stable across frequency scales plus one column with dissimilar connectivity-states for different frequency scales. 

To compare the differences of $\langle \overline{I_{\Delta f}^{sim}}\rangle(k)$ in more detail at the level of every realization of these iterations, \textit{post hoc} right-tailed, signed two-sample Wilcoxon rank-sum tests were evaluated between all possible pairings of $k$s, i.\,e. we tested the nine data points of each $k$ in plot \ref{fig:scaleStabilityIndex}B1 against each other, and did so for every frequency-shuffled realization. Figure \ref{fig:scaleStabilityIndex}C1 shows the number of significant Wilcoxon rank-sum tests at a level of $\alpha = 0.00196$ (Bonferroni-corrected $p$-value for each realization is $p = .0014$). With this procedure, and even though there are few data points being compared, we get a distinct region for $k\in [3;4]$. For these two $k$s, almost all of those Wilcoxon rank-sum tests result in a rejection of the null-hypothesis of equal averages for all compared pairs of $k$ except $k = 2$ (red rows/columns in fig. \ref{fig:scaleStabilityIndex}C1). All remaining pairings result in almost no significant Wilcoxon rank-sum tests (blue areas) except comparisons of $k = 2$ and $k$-means runs with high $k$ (green areas). Thus, by using this method, we get a distinction between scale-stable $k$-means runs and states, which would result in  connectivity-states unstable across  frequency scales. Clusters with $k = 3$ and $k = 4$ show distinct stability across frequency scales compared to other similarity groupings. Since there is one study \cite{Leonardi2014b} also suggesting demeaning of each correlation function in the dFC matrices, we also evaluated our approach on demeaned $\mathbf{X}_i^{S}(t)$ and $\mathbf{X}_i^{I_f}(t)$. This means that from each component of those matrices the temporal mean of those component is subtracted. Afterwards those sets of matrices enter the clustering procedure. Also this modification results in similar patterns (see fig. \ref{fig:scaleStabilityIndex}A4,C4) like the canonical approach (see fig. \ref{fig:scaleStabilityIndex}A1,C1). One deviation to the canonical dataset is the emergence of stability for $k = 6$ and $k = 8$ in comparison to $k = 7$ and $k = 9$. Nevertheless, the main finding of large $\langle \overline{I_{\Delta f}^{sim}}\rangle(k)$ for $k = 3$ and $k = 4$ and the drop-off from $k = 4$ to $k = 5$ still holds and the structure of the connectivity-states is still comparable (fig. \ref{fig:scaleStability_robustnessK4_demeaned}) besides the stabilizing connectivity-state that resembles the average connectivity is missing.

\subsubsection{Theoretical considerations}
Assuming there are $k_{inh}$ connectivity-states inherent to the data and $k$-means identifies those $k_{inh}$ connectivity-states as centroids in each run and also at each frequency scale $f$ perfectly, then the behavior of $\langle \overline{I_{\Delta f}^{sim}}\rangle(k)$ can be deduced in this ideal case. Additionally we assume that the cluster centroids that do not belong to connectivity-states express zero correlation with each other and with the connectivity-states. This considerations imply that the shuffling of frequency scales is redundant, because both non-connectivity-states and connectivity-states always align across frequency scales. Also averaging over different $\Delta f$ is redundant, because each correlation either results in a correlation coefficient of $\rho = 0$ or $\rho = 1$. Therefore any $\Delta f$ represents the sum equally and the sum over $f$ in equation \ref{eq:simOverScales} can be discarded. With the perfect alignment of connectivity-states and non-connectivity-states across scales the different columns in the plot arrays can either be represented by $\rho = 1$ or $\rho = 0$. This means that in this hypothetical case $\langle \overline{I_{\Delta f}^{sim}}\rangle(k)$ reduces to
\begin{equation}
\langle \overline{I_{\Delta f}^{sim}}\rangle(k) =
\begin{cases}
1, &\mbox{if } k \leq k_{inh}\\
\frac{k_{inh}}{k}, &\mbox{otherwise}. 
\end{cases}
\end{equation}
Thus, theoretically, if $k$-means always finds the correct connectivity-states for each $k$ and in each scale $f$, then $\langle \overline{I_{\Delta f}^{sim}}\rangle(k)$ stays constant until the desired number of connectivity-states is reached and then a drop-off with $\frac{k_{inh}}{k}$ occurs.

\subsection{Null-model tests}
We tested the validity of our results by applying our approach to two null-models derived from the original $\mathbf{T}_i(t)$. Time courses from both null-models entered the pipeline at the same level as the original time courses $\mathbf{T}_i(t)$. The $k$-means algorithm was applied with $200$ iterations (example plots can be found in figs. \ref{fig:scaleStability_robustnessK4_shuffled} and \ref{fig:scaleStability_robustnessK4_phaseRand}). The first null-model was introduced by shuffling the standard time courses $\mathbf{T}_i(t)$ for each session $i$ and each RSN $c = 1, \dots, C$, which destroys any temporal structure present in the data. The second null-model was derived from phase-randomizing each RSN time course from $\mathbf{T}_i(t)$. This null-model ensures that frequency relevant structure is preserved in the time courses while destroying the covariance patterns \cite{Allen2012,Prichard1994}. An MEMD yielded more IMFs (fig. \ref{fig:scaleStability_robustnessK4_shuffled}) for the shuffled time courses, since shuffling introduces high frequencies not present beforehand.

\begin{figure}[h!tbp]
\centering
\includegraphics[width = 1\textwidth]{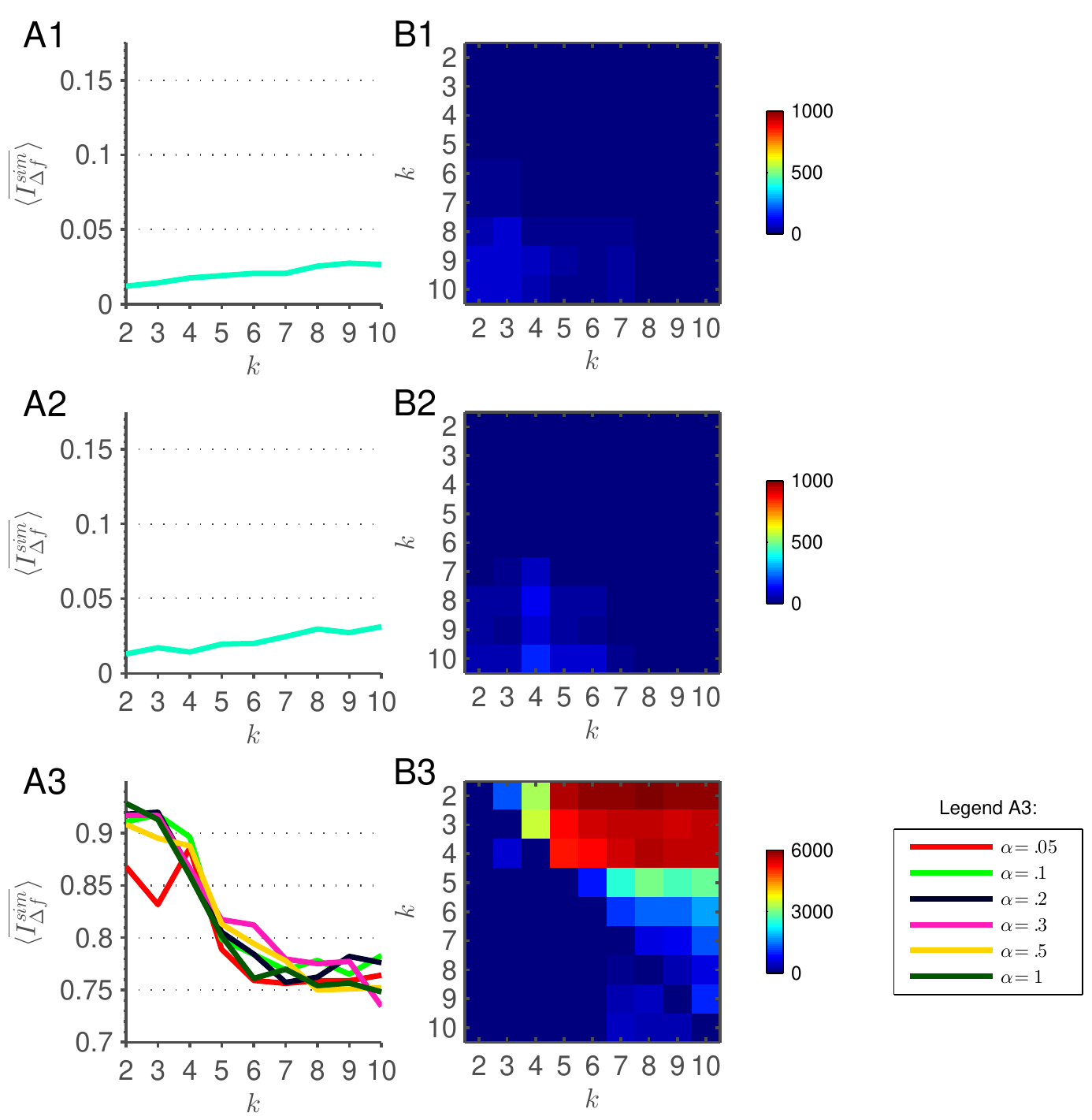}
\caption{This figure depicts the null-model tests of the introduced approach and shows the results of the adjusted window sizes. In panel (A1) $\langle \overline{I_{\Delta f}^{sim}}\rangle(k)$ for shuffled and in panel (A2) for phase-randomized time courses is plotted. Both surrogate time-series show similar behavior with almost vanishing $\langle \overline{I_{\Delta f}^{sim}}\rangle(k)$ for all $k$ and a slightly opposing trend to {\protect fig. \ref{fig:scaleStabilityIndex}A1}. Also the corresponding hypothesis test patterns in panels (B1) and (B2) show very few significant tests and emphasize the opposing trend. In panel (A3) the results of the adjusted window size approach is plotted. The different $\alpha$ values correspond to more or less adjusted window sizes. Panel (B3) depicts the merged hypothesis test patterns over all $\alpha$-values.}
\label{fig:nullModels}
\end{figure}

The results of the null-model tests are shown in figure \ref{fig:nullModels}A1-2,B1-2. It can be seen that both approaches with surrogate time-series result in an almost vanishing scale-stability index (shuffled: $\langle \overline{I_{\Delta f}^{sim}}\rangle(k) < .028$; phase randomized: $\langle \overline{I_{\Delta f}^{sim}}\rangle(k) < .032$) or even show an opposing trend with slightly increasing stability with increasing $k$ (fig. \ref{fig:nullModels}A1-2). The Wilcoxon rank-sum tests result in a small number of significant tests (shuffled: $n^{wilc.}_{max} = 92$; phase randomized: $n^{wilc.}_{max} = 169$) and the pattern in the corresponding panels (fig. \ref{fig:nullModels}B1-2) stresses the mentioned opposing trend.

\subsection{Frequency dependent window size}
We also investigated the dependence of the window size used in the sliding-window approach. The above reported results are extracted using a constant window size of $w(\overline{f}) = \SI{57.6}{\s}$. This size is applied both to the frequency-resolved time courses resulting from MEMD, and the original time courses. To provide a reference of how many periods are covered by this window size in the original time courses, we calculated the fast Fourier transformation of the average spectrum over all used RSNs and sessions. The weighted-average frequency is $\overline{f} \approx \SI{0.055}{\Hz}$ and the number of periods covered by a window with size $w(\overline{f})$ is $\overline{n_T} \approx 3.17$. To adjust the window size to the average instantaneous frequency $f$ of the IMFs, we added

\begin{equation}
\Delta w^{\alpha}(f) = w(\overline{f})\left(\frac{\overline{f}}{f} - 1\right)\cdot \alpha
\label{eq:adjW}
\end{equation}

to $w(\overline{f})$, resulting in an adjusted, frequency dependent window size $w^{\alpha}(f)$. We chose to use $\alpha \in \{0.05, 0.1, 0.2, 0.3, 0.5, 1\}$ resulting in $w^{\alpha}(f)$ as plotted in figure \ref{fig:adjW}. For $\alpha = 0.5 \land f = f_9$ and $\alpha \in \{0.5, 1\} \land f = f_{10}$ $w^{\alpha}(f)$ exceeded $T$. In those extreme cases, window size was chosen to be $w^{\alpha}(f) = T - 1$. For $\alpha = 1$, the frequency dependence resembles the case where the number of periods in a window is equal across all frequency scales (except the mentioned cases) 

\begin{equation}
w^{1}(f) = w(\overline{f})\cdot\frac{\overline{f}}{f}.
\label{eq:adjWmax}
\end{equation}

Afterwards, the sliding window approach was conducted with window size adapted to different frequency scales. For each $\alpha$-value and each frequency scale $f$, $k$-means clustering was applied in analogy to our introduced procedure, which would be represented by $\alpha = 0$. For each $\alpha$-value, $\langle \overline{I_{\Delta f}^{sim}}\rangle(k)$ was calculated and plotted against $k$ (fig. \ref{fig:nullModels}A3). Also with variable window size a comparable pattern emerges with larger drop-offs from $k = 4$ to $k = 5$. For each $\alpha$, we also evaluated the same hypothesis test procedure introduced above. The results are merged in figure \ref{fig:nullModels}B3, in which the results of the Wilcoxon rank-sum tests are summed up for each possible comparison over all $\alpha$-values. The same significance level as in the approach with constant window size is used.

\subsection{Simulated dynamic functional connectivity}
\label{sec:simulation}
\begin{figure}[h!tbp]
\centering
\includegraphics[width = 1\textwidth]{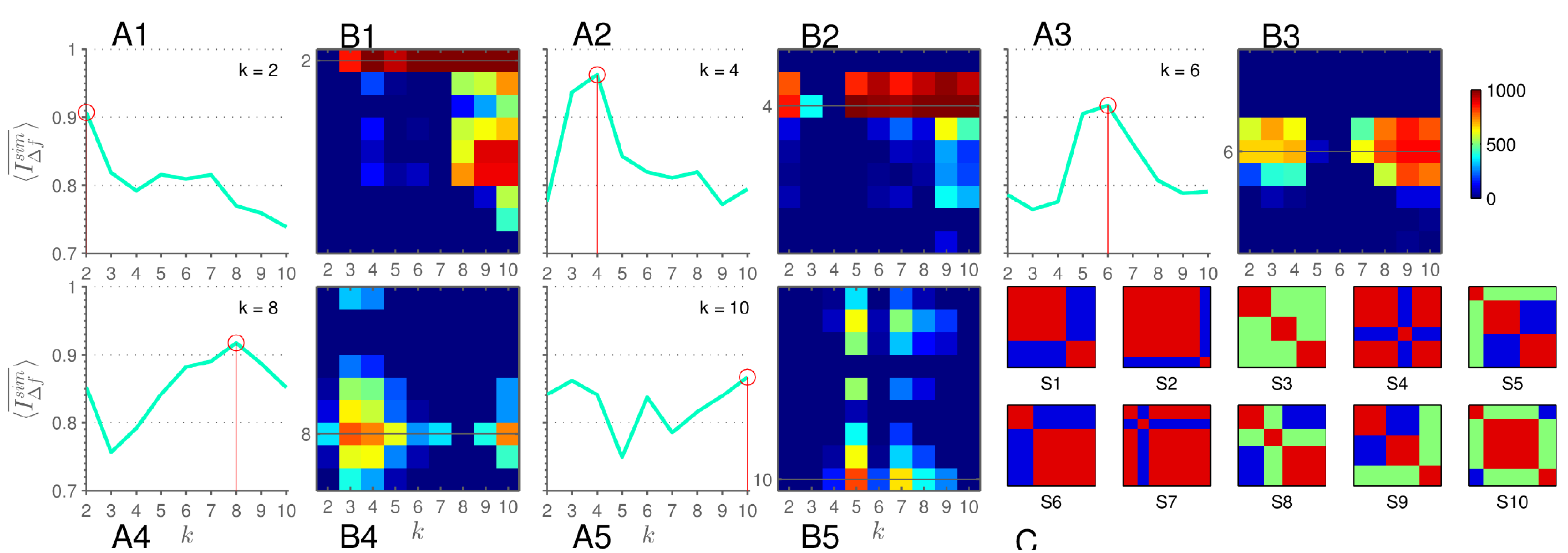}
\caption{This figure depicts the simulation results using the SimTB toolbox for simulating rs-fMRI time-series. We simulated dFC traversing $k_{sim}\in \{2, 4, 6, 8, 10\}$ artificial connectivity-states shown in panel (C). Panels (A1-5) show the results of $\langle \overline{I_{\Delta f}^{sim}}\rangle(k)$ and panels (B1-5) the corresponding pairwise signed Wilcoxon ranks-sum tests for increasing $k_{sim}$ respectively. The pattern in panel A2 and B2 is similar to the corresponding in figure \ref{fig:scaleStabilityIndex}. In general the combination of $\langle \overline{I_{\Delta f}^{sim}}\rangle(k)$ and the hypothesis tests of  $\overline{I_{\Delta f}^{sim}}(k)$ seem to offer a way to distinguish between $k$-means runs with a distinct number of $k_{inh}$ connectivity-states.(A,B1: $p_u = .8, a_u = .8, p_s = 1$; A,B2: $p_u = .8, a_u = .6, p_s = 8$; A,B3: $p_u = .6, a_u = .8, p_s = 1$; A,B4: $p_u = .8, a_u = .8, p_s = 1$; A,B5: $p_u = .8, a_u = .6, p_s = .8$; all parameters sets are characterized by smaller values in the noise terms compared to the coherence term.)}
\label{fig:simSummary}
\end{figure}

In order to validate our approach and strengthen our results, we applied our method to simulated data from the toolbox \textit{SimTB} \cite{Erhardt2012}, which enables us to simulate dFC traversing a predefined number of connectivity-states $k_{sim}$. To do that we modified the script\footnote{http://mialab.mrn.org/software/simtb/docs/create\_toysimulation.m} offered in the course of the study of \cite{Allen2012}. We simulated $24$ RSN time courses with a length of $M = 1200$ time points and a $\mathrm{TR} = \SI{.7}{\s}$. No Gaussian noise was added to the time courses, since the real time-series were denoised before entering MEMD. The number of simulated subjects and sessions was chosen so that in each simulation run with a particular number of artificial connectivity-states the number of occurrences of each state was $n_{occ.} = 96$. The sequence of connectivity-states was randomized, but with the constraint that consecutive intervals always traverse different connectivity-states. Thus, the transition between states was randomized. The duration of each connectivity-state was chosen to be $\Delta T = 150$ TRs. We chose the numbers of connectivity-states to be simulated as $k_{sim} \in \{2, 4, 6, 8, 10\}$, the prototypes of which can be found in figure \ref{fig:simSummary}C. There are three main parameters that can be adjusted -- the probability of a unique event $p_u$ and its amplitude $a_u$ in relation to the amplitude of the coherent, or rather, state event with the occurence probability $p_u$. We probed this three dimensional parameter space, but an exhaustive parameter sweep was not possible due to the high computational demand. We simulated all possible combinations of the values $\{.6, .8, 1\}$. Additionally, we calculated the iMMSE for our rs-fMRI data and for each of the $13$ simulation runs (fig. \ref{fig:iMMSE_sim}) in order to choose the best parameter set in terms of complexity.

Additionally, we conduct complexity analysis in terms of iMMSE of the $400$ rs-fMRI sessions used in this study and for each of the $13$ simulation runs (fig. \ref{fig:iMMSE_sim}) in order to choose the best parameter set in terms of complexity. We used literature-based values for the parameters of lag-length $\tau = 1$, threshold $r = .15\sigma$, and length of the template vector $m = 2$ \cite{Ahmed2012}, since iMMSE analysis has never been done so far on rs-fMRI data. The structure of the iMMSE curve shown in figure \ref{fig:iMMSE_sim} is similar to the finding of \cite{McDonough2014}, who used univariate multiscale sample entropy based on coarse graining. The similarity is strengthened by the fact derived by \cite{Ahmed2012} that the dependence of scales $\varepsilon$ from multiscale sample entropy approaches using coarse graining and the iMMSE method $f$ relying on IMFs is $\varepsilon \approx 2^{f-1}$. The simulated time courses also show similar complexity structure, but none of the runs could fit the iMMSE of rs-fMRI data. Therefore, we show a selection of parameter sets in figure \ref{fig:simSummary} as a proof-of-principle. The data of all the other simulation runs can be found in our data repository.

$\langle \overline{I_{\Delta f}^{sim}}\rangle(k)$ peaked for $k_{sim}\in \{2, 4, 6, 8, 10\}$ at the corresponding $k$ (fig. \ref{fig:simSummary}A1-5). The hypothesis test plots reveal distinctive patterns between $k_{sim}$ runs. Also for $k_{sim}\in \{2, 4, 6, 8, 10\}$ the patterns result in the detection of the right number of connectivity-states (fig. \ref{fig:simSummary}B1-5). Considering the theoretical considerations above the expected drop-off also shows up comparable to the corresponding plots of the rs-fMRI data (fig. \ref{fig:scaleStabilityIndex}A*).

\subsection{Alternative frequency scale definitions}
MEMD also has a unique feature of extracting instantaneous frequency and amplitude information for each scale introduced above. By exploiting this we create frequency scales with normalized amplitude (eq. \ref{eq:uniformAmp}). Thus, we are able to conduct frdFC analysis on scales without any amplitude information. Our approach applied on this type of scales converges in figure \ref{fig:scaleStabilityIndex}A2-C2. A plateau results for $k \leq 4$ with a drop-off behavior comparable to the canonical approach. Taking the theoretical considerations of $\langle \overline{I_{\Delta f}^{sim}}\rangle(k)$ into account this type of scale shows the expected plateau for $k \leq k_{inh}$ where the data inherent connectivity-states would also be estimated as $k_{inh} = 4$. But we want to be careful with claiming a distinct plateau from $k = 2$ to $k = 4$, since in the comparison of liberal, intermediate, and conservative datasets, the case $k = 2$ was less stable in terms of $\langle \overline{I_{\Delta f}^{sim}}\rangle(k)$ and the corresponding hypothesis test plots except from the stable findings for $k = 3$ and $k = 4$. Nevertheless, the plateau at $k = 3$ and $k = 4$ seems to be a very robust finding for our rs-fMRI data.

As another alternative frequency scale definition we investigate cumulative scales. In order to conduct iMMSE analysis the scales offered by MEMD have to be cumulated to represent coarse grained versions of the original time courses (eq. \ref{eq:cumIMFs}). Those time courses $\mathbf{U}_{cum}^f(t)$ can also be used as a substrate to enter frdFC. We conducted our introduced approach on $\mathbf{U}_{cum}^f(t)$, which results in a $\langle \overline{I_{\Delta f}^{sim}}\rangle(k)$ shown in figure \ref{fig:scaleStabilityIndex}A3, which peaked for $k = 3$ but lacking the clear separated region found in figures \ref{fig:scaleStabilityIndex}A1 and A2. In general, the similarity of connectivity-states over those scales is more pronounced than in the other frequency scales lacking the steep drop (compare fig. \ref{fig:scaleStabilityIndex}B1-3)

\subsection{Filter-banks}
In a \textit{post hoc} analysis we apply Butterworth filter-banks both on the rs-fMRI and simulated time courses. We find that a filter-bank emulating bands with similar bandwidths like the frequency scales resulting from MEMD cannot be realized in a sensible manner. The frequency scales represented by IMF indices $f_7$ to $f_{10}$ have a very narrow bandwidth, lie close together, and are represented by frequencies very close to zero (fig. \ref{fig:exampleIMFs_PSD_boxPl}B). Therefore, we switch to  filter-banks with equidistant bands. All orders of the used filters are chosen to result in stable pole behavior. We followed a more canonical and a more adjusted way of constructing the filter-banks. In the canonical way, the order of the filters was chosen to be constant for filter-banks of a certain number of bands. Hence, we chose order 10 for $n_{bands} = \{5,8\}$ bands, order 8 for  $n_{bands} = \{10,12\}$ bands, and order 6 for  $n_{bands} = 15$ bands. In the adjusted way, we design the filters with the maximum possible order barely resulting in stable filter behavior adjusted for each band in each filter-bank. This results in varying filter orders over different bands. The two filter-bank designs can also be considered as a conservative and a liberal design, respectively. For both approaches, the lowest bands are realized as low-pass filters. Since the preprocessing of the rs-fMRI time courses includes a low-pass filter with a cut-off frequency of $\SI{.15}{\Hz}$, we partition the interval $[\SI{0}{\Hz}; \SI{.15}{\Hz}]$ in bands each with bandwidth $\Delta \nu = \frac{\SI{.15}{\Hz}}{n_{bands}}$. We evaluate our approach on the resulting time-series on  $n_{bands}$ different frequency scales.

The pattern shown in figure \ref{fig:scaleStabilityIndex}A,B,C5 is comparable to the ones shown in panels A,B,C1,2,4. We find for the filter-banks with constant order \ref{fig:scaleStabilityIndex_filterBanks_realTCs_new} that the results using 10 and 12 bands best resemble the other, non-filter-bank findings. For 8 and 15 bands the fourth connectivity-state becomes instable. The filter-banks with the maximum possible order of the filters results in a less stable picture \ref{fig:scaleStabilityIndex_filterBanks_realTCs_new2}. Here, 8 bands show the drop-off, which is less robustly found for the other numbers of bands. In general, 5 bands show less distinct patterns and the hypothesis test plots become redundant. The same logic is applied to the simulation run with $k_{sim} = 4$ simulated connectivity-states. For both approaches (figs. \ref{fig:scaleStabilityIndex_filterBanks_simTCs_new} and \ref{fig:scaleStabilityIndex_filterBanks_simTCs_new2}) we find that the main pattern (fig. \ref{fig:simSummary}A,B2) is evident for all filter-bank realizations besides 5 bands. When increasing the number of bands the main pattern becomes more evident.

\subsection{Data repository}
We provide a data repository of our frequency-resolved time-series.\footnote{http://doi.org/10.5283/epub.32642} In this repository we include the canonical scales (fig. \ref{fig:scaleStabilityIndex}A1,B1,C1), uniform-amplitude scales (fig. \ref{fig:scaleStabilityIndex}A2,B2,C2), cumulative scales (fig. \ref{fig:scaleStabilityIndex}A3,B3,C3), the time-series resulting from the filter-bank procedures both for constant and adjusted filter order (figs. \ref{fig:scaleStabilityIndex_filterBanks_realTCs_new},\ref{fig:scaleStabilityIndex_filterBanks_realTCs_new2},\ref{fig:scaleStabilityIndex_filterBanks_simTCs_new},\ref{fig:scaleStabilityIndex_filterBanks_simTCs_new2}), and the MEMD decomposed time-series from the simulation runs elaborated on in section \ref{sec:simulation} plus the not shown simulations (all main effects and interactions). In addition to the decompositions we also offer the original, non-decomposed time courses with the full preprocessing level where applicable (not for simulations). The dFC matrices are stored on local servers and can be retrieved on demand.

\section{Discussion}
\subsection{General statement}
We applied MEMD to complete time courses of RSNs resulting from a gICA analysis, i.~e. before the sliding-window procedure was employed to perform a dFC analysis. We showed that MEMD offers the possibility of comparing IMFs over sessions and RSNs, since it aligns the modes accurately. This results in time courses that can be compared within each IMF-level across all sessions and RSNs, opening up a way to perform a dFC analysis on different frequency scales. This suggested a frdFC analysis and revealed a  similarity of connectivity-states which was robust across all inherent frequency scales, even the lowest frequencies contained in the time-series. The dependence of scale-stability of connectivity-states on the number of extracted clusters is a novel finding and could be related to existing literature describing scale-invariance as a characteristic feature of rs-fMRI. We compared our results to several null-models and this comparison convincingly corroborated our claims. Additionally we explored two more frequency scales in terms of frdFC that completes our aim to offer a baseline for frdFC research by means of MEMD.

The combination of the properties of accurate mode alignment and feature extraction in terms of resulting IMFs suggests MEMD as a unique tool for performing a frdFC analysis. However, comparison to other frequency-analysis techniques like Fourier or wavelet transforms is not possible directly. Remember that MEMD operates in the time domain and never transforms the signal to a conjugate space. Further it expands the signal in a set of modes which are adaptively deduced from the data rather than expanding it into a basis function set pre-defined and possibly inappropriate. Thus MEMD offers a way to apply sliding-window techniques to intrinsic modes with problem specific frequency scales.

When it comes to apply MEMD, time courses are decomposed into a locally orthogonal set of IMFs that -- when summed up -- result in the original time-series. Investigating frdFC by means of MEMD yields two conflicting considerations to be taken into account. On the one hand, since time courses are analyzed at different time scales and the related local frequencies are decreasing with increasing IMF index, adjusting window size by equation \ref{eq:adjWmax} is an obvious adaptation of the approach. On the other hand, the representation of the original time course by the superposition of all IMFs prefers the case with constant window size where $\alpha = 0$ in equation \ref{eq:adjW}. This assures that, within any given time window, time samples of different modes correspond accurately to each other across all modes, i.\,e., each time point has its exact partner for each IMF index. As a consequence, choosing an adaptive window size yields correlation matrices which are less comparable across different frequency scales, since for modes with higher IMF-indices, the time window encompasses time samples which do not belong to corresponding time windows in modes with lower IMF-indices. However, choosing $\alpha = 0$ neglects the increasing period of intrinsic local oscillations with increasing IMF index. Those two lines of thought have to be kept in mind when performing frdFC analysis by means of MEMD. Yet another consideration favors the approach with constant window size when it comes to apply $k$-means clustering. Adapting window size to the period of local oscillations in different modes also leads to varying numbers of data points in the clusters if $k$-means is applied at different frequency scales. In the extreme case with $\alpha = 1$, $k$-means runs with $n_{f_1} = 461600$ and $n_{f_{10}} = 400$ data points are compared. Statistically, this severe imbalance offers the algorithm the possibility to extract a larger number of centroids for IMF index $f_1$ and a vastly reduced space of possible centroids for $f_{10}$. With these considerations in mind, the approach with constant window size seems to be preferable over the one with time scale adapted window sizes. Furthermore, the striking similarity of connectivity-states across frequency scales for $k = 4$ in the constant window case lends credit to prefer this approach over the adaptive one. But nevertheless, when considering both approaches equivalently, our results still consistently show a high $\langle \overline{I_{\Delta f}^{sim}}\rangle(k)$ for $k \leq 4$ and above this value $\langle \overline{I_{\Delta f}^{sim}}\rangle(k)$ drops. The simulated dFC traversing artificial connectivity-states (fig. \ref{fig:simSummary})  strengthen the result of $k = 4$ inherent scale-invariant states in the data. The peak of $\langle \overline{I_{\Delta f}^{sim}}\rangle(k)$ is sharper for $k_{sim} = 4$ in the simulated case (fig. \ref{fig:simSummary}B2), which can be expected in such an idealization. Moreover, the pattern shows a similar region for $k > 4$ on the abscissa and $k = [3, 4]$ on the ordinate like it was found in real data (fig. \ref{fig:scaleStabilityIndex}B) that is distinct from the other simulated cases.
 
\subsection{Scale-stability and its possible benefits for connectivity-state extraction}
In the course of this study we also looked at elbow-criteria after $k$-means runs as a selection criterion for the proper number of extracted connectivity-states. Unfortunately, no clear elbow-criterion could ever be established. Using within-cluster similarity as a measure of choice, clear elbows are rarely evident. Therefore selection of a certain number of extracted clusters would have been always subjective.

Regarding this fact, and our results concerning varying scale-stability of connectivity-states across frequency scales and model order $k$, we rather suggest to consider scale-stability $\langle \overline{I_{\Delta f}^{sim}}\rangle(k)$ as a proper measure to infer model order. The latter could be identified as the number of clusters/connectivity-states which show highest scale stability across the relevant frequency scales. Our study shows that $k = 4$ connectivity-states deem most appropriate for the data that has been analyzed. This optimization for scale-stability is in line with literature reports which suggest scale-invariance as an inherent feature of rs-fMRI. The large number of significant Wilcoxon rank-sum test results for $k\in[3;4]$ of $I_{\Delta f}^{sim}(k)$ can be interpreted as indicating these $k$-means runs as the ones with, in terms of $\langle \overline{I_{\Delta f}^{sim}}\rangle(k)$, a significantly more stable structure of connectivity-states across all inherent frequency scales  compared to other model orders $k$. In \cite{Kitzbichler2009}, the authors argue that phase-synchrony is an important feature for network formation at all frequencies in resting-state conditions. Since the extraction of connectivity-states from dynamically changing correlation matrices implies mutual coherence, a similar argument holds here. Thus this argument is also in favor of identifying model order with the number $k$ of connectivity-states with highest scale-stability across all inherent frequency scales.

By evaluating hypothesis-tests resulting in plot \ref{fig:scaleStabilityIndex}B1, model orders $k \in [3;4]$ indicate high scale-stability, while for all other model orders the corresponding connectivity-states are not scale-invariant across different inherent frequency regimes as is corroborated by a much smaller stability measure $\langle \overline{I_{\Delta f}^{sim}}\rangle(k)$. The latter thus breaks up into two regimes with highly stable and scale-invariant connectivity-states for model orders $k \in[3;4]$ and structurally very fragile connectivity-states for model orders $k < 2$ and $k = 5, \dots, 10$. Especially for model orders $k > 4$ the stability measure $\langle \overline{I_{\Delta f}^{sim}}\rangle(k)$  drops dramatically. Thus, in terms of maximizing the information to be extracted from the system by minimizing trade-offs, our results favor a model order $k = 4$, i.\,e., four scale-invariant and structurally stable connectivity-states. Generally, by combining our scale-stability index with an hypothesis testing procedure, the simulation and rs-fMRI results suggest these measures suitable for differentiating between data with different numbers of inherent connectivity-states. This procedure results in a quality criterion for the estimation of the number of data inherent connectivity-states, which is based on the feature scale-invariance common to rs-fMRI parameters, differing from the statistical quality criterion introduced in the course of the study of \cite{Leonardi2014b}.

The main finding of a high level of stability for $k = 3$ and $k = 4$ and the drop-off from $k = 4$ to $k = 5$ is robust over several modifications. The uniform-amplitude scales show (see fig. \ref{fig:scaleStabilityIndex}A2,C2) almost the same pattern like the canonical frequency scales (see fig. \ref{fig:scaleStabilityIndex}A1,C1). The difference found for $k = 2$ could point to an advantage of the uniform scales over the canonical scales concerning the theoretical considerations we deduced for $\langle \overline{I_{\Delta f}^{sim}}\rangle(k)$ namely the plateau from $k = 2$ to $k = 4$ that would be complemented by the high value of $\langle \overline{I_{\Delta f}^{sim}}\rangle(k)$ for $k = 2$. In our stability checks with different numbers of ICs interpreted as RSNs (liberal, intermediate, conservative set of RSNs introduced above) we found that for $k = 2$ $\langle \overline{I_{\Delta f}^{sim}}\rangle(k)$ is less stable than for other $k$. Therefore, even though this interpretation would prefer the uniform-amplitude scales, we want to mention it with caution. The cumulative scales lead to a less clearer pattern in the hypothesis test plot (fig. \ref{fig:scaleStabilityIndex}C3). The drop-off from $k = 4$ to $k = 5$ of $\langle \overline{I_{\Delta f}^{sim}}\rangle(k)$ (fig. \ref{fig:scaleStabilityIndex}A3) is statistically not as pronounced as in the latter two frequency scales. Intriguingly the recently suggested component-wise temporal demeaning of dFC matrices \cite{Leonardi2014b} yields similar results in terms of stability for $k = 3$ and $k = 4$. \cite{Leonardi2014b} used a considerably smaller sample of sessions compared to our dataset. We suggest that component-wise temporal demeaning seems to be more crucial for studies with smaller cohorts than large population studies and that its effects average out with increased number of used sessions.  We suppose that for smaller cohort studies demeaning seems to be a crucial aspect \cite{Leonardi2014b}, but for large population studies the effect of demeaning seems to diminish in terms of feature extraction by means of $k$-means. Also the simulations conducted (fig. \ref{fig:simSummary}) point to a benefit in scale-stability considerations when it comes to inferring the underlying number of connectivity-states in a dataset. For each $k_{sim}$ a distinct pattern in $\langle \overline{I_{\Delta f}^{sim}}\rangle(k)$ and the corresponding hypothesis test plot emerged. When considering the scale-stability findings of MEMD, filter-banks, simulated and rs-fMRI data together, then we have accumulated evidence that data-inherent connectivity-states emerge in a stable manner over several scales.

\subsection{Filter-banks, MEMD, and self-similarity}
Analyzing expectably self-similar time-series by means of EMD is a natural way of approaching such data \cite{Flandrin2004}. Exploring the frequency structure in more detail via filter-banks revealed striking robustness of the scale-invariance feature concerning connectivity-states. Especially, in the case of simulated time courses cycling through $k_{sim} = 4$ artificial connectivity-states, where the scale-invariance of those states holds for different numbers of bands, i.\,e., different frequency resolutions. For rs-fMRI time-series the variation in the number of bands and with it the variation in bandwidth yields less robust results, especially in the fourth connectivity-state. For real data it seems that the design of the filter-bank is more crucial than for simulated data. Nevertheless, we conclude that real and even simulated rs-fMRI time-series cycling through connectivity-states imprint their coherence not only on very distinct frequencies, but rather on a much broader spectrum. MEMD in turn seems to be capable of finding bands in a data-driven manner that consist enough information to reveal the scale-invariance of connectivity-states. Nevertheless, also filter-banks posses this capability, but may be more sensitive to fine tuning parameters.

We want to stress the fact that the pattern in the simulated (fig. \ref{fig:simSummary}B2) and rs-fMRI data (fig. \ref{fig:scaleStabilityIndex}C1) are strikingly similar. This pattern can also be found in most of the parameter combinations of the simulated data. Comparing the filter-bank findings from the simulated data (figs. \ref{fig:scaleStabilityIndex_filterBanks_simTCs_new},\ref{fig:scaleStabilityIndex_filterBanks_simTCs_new2}) with those two MEMD findings, the main pattern always holds and can be carved out when neatly designing the filter-bank. Considering the rs-fMRI data it seems that the design of the filter-bank has to be more precisely chosen than for artificial data, since the patterns shown in figures \ref{fig:scaleStabilityIndex_filterBanks_realTCs_new} and \ref{fig:scaleStabilityIndex_filterBanks_realTCs_new2} vary slightly depending on the chosen filter-bank. This finding is not surprising, since in real data more noise-types should be present in the time courses than can be simulated by the unique events introduced to the artificial time-series. Since, the finding of a scale-stability drop-off from $k = 4$ to $k = 5$ is this common in our studies, we consider it as a benchmark pattern concerning our used approaches and scale-definitions. Assuming this pattern being the desired outcome of the stability analysis, we find that the results from the filter-banks are susceptible to parameter tuning of the filter. Moreover, the mere difference in complexity of the simulated and rs-fMRI time-series found in our iMMSE investigations shows that simulation cannot completely emulate the real data. We argue that this fact is also reflected in the comparison of simulated and rs-fMRI filter-bank data. Nevertheless, MEMD seems to be capable of finding the bands, which result in similar patterns in simulated and real data.

\subsection{Limitations and future directions}
The time scales of extracted IMFs have also to be discussed. Recent research showed that fluctuations in cortical activity have physiological foundations even in the infra-slow frequency range \cite{Hiltunen2014,Pan2013}. This means that intrinsic modes with indices up to $f_7$ are still of physiological origin. For even slower IMFs, corresponding to large periods or, equivalently, low instantaneous frequencies, a physiological interpretation is not immediately obvious and needs further investigations. In this respect it is worth mentioning that even on long (low) time (frequency) scales, the structure of the four connectivity-states, which we were able to identify, is still preserved. Hence, we are not dealing with artifacts on these time scales. Thus the question arises: why is it possible to literally define connectivity-states exploring the slowest fluctuations, which result from the decomposition of cortical activity and exploiting the limited information available from them? We claim that our findings corroborate self-similarity and scale-freeness, which indeed has been found in rs-fMRI \cite{Kitzbichler2009,Eguiluz2005,Fraiman2009,Tagliazucchi2012}, which demonstrates the emergence of similar structures over a wide frequency range.

Because the multivariate extension of EMD applied in this study is computationally very costly, we had to restrict our decomposition to 30 ICs. As mentioned in the methods section, extracting this number of ICs is still valid according available literature results. However, with higher-dimensional decompositions we would expect that cerebellar and subcortical ICs would have a much larger signal-to-noise ratio. But as our simulations with conservative, intermediate, and liberal sets of RSNs showed very robust and similar results, we are confident that using higher dimensional decompositions would merely strengthen our results.

Besides those limitations our method is not only able to extend the standard dFC to an frdFC approach, rather it also shows that merely looking at the separation of frequency scales by MEMD already reveals benefits in investigating brain networks. While brain graphs are traditionally based on thresholding correlation matrices, our approach offers a new way of simultaneously investigating the dynamics of such brain connectivity networks on different frequency scales. This is achieved by thresholding the static or dynamic correlation matrices calculated from time courses separated into different intrinsic modes. Thus, for each subject or session not only one static brain graph or one set of dynamic brain graphs can be constructed, rather brain connectomes can be resolved at various inherent frequency scales. This offers the possibility to adapt brain graphs to inherent and problem-specific time scales, and investigate graph theoretical properties like small-worldness, corresponding to different inherent dynamics simultaneously. This means that the emerging diagnostic values of functional connectivity in diseases like Alzheimer's disease and related dementias, schizophrenia or Parkinson's disease could benefit from a higher diagnostic sensitivity by looking at functional connectivity -- and with it brain graphs -- on different time scales. Also methods investigating cognitive states by relying on brain graphs \cite{Cribben2012b} can adapt additional degrees of freedom by applying certain aspects of our approach.

By introducing an MEMD decomposition to develop a frdFC analysis, we also established a link between complexity research and dFC by exploring cumulative scales commonly used in iMMSE studies. For future research it would be interesting to delve into this subject even further and gain additional insight in the changing SE over time. Further studies could use iMMSE as a dynamic measure, which might reveal connectivity-state transitions by sharp changes in complexity between consecutive windows.

In general we conducted our study on a very large dataset consisting of $400$ sessions with $1115$ windows each. Running clustering algorithms on such a huge dataset can only account for very general features inherent to the data of all subjects -- even if demeaning of the single correlation functions has been conducted. Such a dataset is suitable for a baseline study like ours, but it has to be replicated for smaller studies with a specific psychological hypothesis, if connectivity-states can also be extracted in a robust manner over scales, if the stability over scales also peaks at $k = 4$ or if scale-stability breaks down. If the latter turns out to be true, complexity-loss theory for systems under stress (represented by the stimulated brain) \cite{Ahmed2011,Goldberger2002} would be a plausible explanation.

Further studies could investigate modifications of our approach to apply MEMD on higher dimensional data. Since MEMD is computationally very costly with increasing number of channels, it would be desirable to develop a method for applying it onto finer parcellated cortices. Further studies could apply PCA on the dataset reducing dimensionality in the spatial domain. Then, on those reduced datasets, MEMD can be evaluated and the resulting PCs can be backreconstructed into the original dimension, but now frequency-resolved. Furthermore, future studies can delve into designing filter-banks and explore e.\,g. the feasible frequency resolution. Our provided repository of frdFC time courses is a suitable baseline for this purpose.

Another intriguing finding is that even in simulated time courses traversing artificial connectivity-states scale-stability is an inherent feature. Since the generating algorithms of those simulations are known, it could be possible to look for a deeper mathematical understanding of the scale-stability emerging from those operations. Thus, further studies could investigate the theoretical framework of scale-stability in the context of dFC and connectivity-states.

\subsection{Conclusion}
Our study offers a baseline for further frdFC research. We introduced frdFC by means of MEMD and explored three types of frequency scales discussed in literature. Intriguingly, our frequency resolution of dFC revealed robust scale-stability of connectivity-states over a wide range of inherent time (frequency) scales. This could be achieved by applying MEMD, which thus seems to be a suitable tool for a frequency-resolved level of dFC analysis. Furthermore, by decomposing time courses of ICs at an early stage of the analysis protocol, we were able to gain deeper insight into the behavior of connectivity-states, which otherwise could not be revealed. Thus our results suggest connectivity-state as a useful concept for cognitive studies and confirm that they are a stable construct, well adapted to a wide range of inherent and problem-specific time scales. Scale-stability was shown to offer a proper means to estimate the underlying model order, thus revealing the proper number of inherent connectivity-states extracted by clustering approaches and structurally stable across inherent dynamical processes on different time scales. \textit{Post hoc} filter-bank studies show that scale-invariance of connectivity-states is a robust feature over more frequency-scales than offered by MEMD. But choosing an ill-shaped filter-bank can result in a mis-detection of the number of data-inherent states. Since MEMD is a data-driven approache, no fine-tuning is necessary for this purpose. Our novel approach to spectrally resolved dFC offers plenty of new degrees of freedom, in which pathological and healthy cognition may be distinguished.

To sum up more specifically (i) our simulated results point to scale-invariance of connectivity-states as being a valid quality criterion for model order selection and (ii) this finding encourages to look for a mathematically deeper understanding of the concept connectivity-state, since in simulated data the generating procedure is known. (iii) Finding similar scale-invariance structure in simulated and rs-fMRI connectivity-states for $k = 4$ is a strong hint towards this number of connectivity-states representing data inherent states. (iv) Our analyses show that scale-stability in both the simulated and rs-fMRI time courses is a robust and strong feature. (v) Choosing the natural way of detecting self-similar structures by using a data-driven approach like MEMD led to further frequency-resolution analyses via filter-banks, which confirmed self-similarity as a data-inherent feature detectable by frequency decompositions. Comprehending all findings yields the conclusion that connectivity-states as a concept are not just a multivariate, but also a multiscale entity.

\begin{appendix}
\renewcommand \thesection A
\section{Appendix}
\subsection{Ordering algorithm}
\begin{algorithm}
\caption{Ordering states over different frequency scales}
\begin{algorithmic}
\REQUIRE $\mathbf{X}_j^{\mbox{I}_{f}}$ correlation matrix of scale $f$ and column $j$
\FOR{nIter = 1 \TO 1000}
\STATE{1) Shuffle scales}
\STATE 2) Check, if unique assignment of states from scale $f + 1$ to the average of the preceding scales is possible by 
\[l_j = \arg\left\{\max_{\forall l}\left[\mbox{corr}\left(\overline{\mathbf{X}}_j^{I_{1\dots f}}, \mathbf{X}^{f + 1}_l\right)\right]\right\},\,\forall j\]
\STATE{3) Assign $l_j$ to corresponding $j, \forall j$ with $l_m \neq l_n, \forall m,n$}
\STATE{4) Get all states with $l_m = l_n$ and assign $l_m$ with maximum correlation coefficient, go to step 2) with remaining unassigned states}
\STATE{5) After all states are assigned go to step 2) and increase scale index $f$ by one}
\STATE{6) After final scale is aligned, save configuration and go to step 1)}
\ENDFOR
\STATE{7) Calculate average scale-stability}
\end{algorithmic}
\label{alg:orderUnique}
\end{algorithm}
The algorithm checks if the average of connectivity-states of the first to the $f$-th scale $\overline{\mathbf{X}}_j^{\mbox{I}_{1\dots f}}$ of column $j$ can be uniquely assigned to one of the connectivity-states of the $f + 1$-th IMF scale $\mathbf{X}_j^{\mbox{I}_{f + 1}}$ by correlation. Each matrix of the $f + 1$-th scale is correlated with each mean of the connectivity-state matrices of the first $f$ scales. The criterion for unique assignment is that each state of the $f + 1$-th scale has its highest correlation coefficient on different columns. If all states of scale $f + 1$ can be uniquely assigned to one column, the algorithm goes on and compares the states of the next scale with the average of the preceding scales and so on. Taking the mean of preceding scales assures that ordering of states does not only rely on one scale, which would be outlier sensitive. If there are states, which cannot be uniquely assigned, i.\,e., which correlate highest on the same column, then the algorithm correlates all those states with columns again, which do not have any assigned partner yet. This is done until all states are uniquely assigned to one column. Before newly correlating the non-unique states the algorithm assigns the state of this set and the column with the highest correlation coefficient. Since averaging over frequency scales is done and we want to have columns with highest similarity over frequency scales, we initiate the above mentioned algorithm 1000 times shuffling scales with each iteration step. Every shuffled initialization has a unique combination of scales. Shuffling is done to avoid any sequence effects. Afterwards, for calculation of simuliarity measure over scales, the average over those realizations is computed. The algorithm then results in connectivity-state assignments with highest similarity of states over frequency scales by using a simple and objective ordering approach. For plot-arrays (figs. \ref{fig:scaleStability_robustnessK4}, \ref{fig:scaleStability_robustnessK2}-\ref{fig:scaleStability_robustnessK10}) and figure \ref{fig:scaleStabilityIndex}B the non-shuffled configurations were used and the connectivity-states from the standard dFC are implemented in the mean $\overline{\mathbf{X}}_j^{\mbox{I}_{1\dots f}}$ to align best columns also concerning similarity with the standard connectivity-states.

Since the correlation values of higher-index frequency scales are strongly pronounced in comparison to lower-index frequency scales, taking the mean $\overline{\mathbf{X}}_j^{\mbox{I}_{1\dots f}}$ could bias the ordering procedure because of weighting higher-index frequency scales in this average more than lower-index frequency scales. To test this concern we normalized all $\mathbf{X}_j^{\mbox{I}_{f}}$ to the range $[-1;1]$ by dividing the positive correlation values by the maximum value and the negative ones by the absolute value of the minimum. This mitigates the discrepancy of lower- and higher-index frequency scales when it comes to ordering. We found that changes of $\langle \overline{I_{\Delta f}^{sim}}\rangle(k)$ take place in the fourth decimal for all $k$. Therefore we consider $\langle \overline{I_{\Delta f}^{sim}}\rangle(k)$ robust concerning this modification.
\end{appendix}

\section*{Disclosure/Conflict-of-Interest Statement}
The authors declare that the research was conducted in the absence of any commercial or financial relationships that could be construed as a potential conflict of interest.

\section*{Author Contributions}
MG designed the approach and did the simulations and analyses. The paper was written by MG, AMT, MWG, and EWL. The study was supervised by MWG and EWL. EWL also contributed to study design.

\section*{Acknowledgments}
Data were provided by the Human Connectome Project, WU-Minn Consortium (Principal Investigators: David Van Essen and Kamil Ugurbil; 1U54MH091657) funded by the 16 NIH Institutes and Centers that support the NIH Blueprint for Neuroscience Research; and by the McDonnell Center for Systems Neuroscience at Washington University. One of the authors (MG) thanks P. Keck and T. Geigenfeind for helpful contributions, and Dr. M. B\"ock for fruitful discussions. The authors declare no conflict of interest.

\bibliographystyle{plain}
\bibliography{library}
\newpage

\thispagestyle{empty}
\clearpage

\thispagestyle{empty}
\clearpage

\thispagestyle{empty}
\clearpage

\thispagestyle{empty}
\clearpage

\thispagestyle{empty}
\clearpage

\thispagestyle{empty}

\clearpage
\section{Supplementary material}
\subsection{Adjustment of GIFT for frequency resolved dFNC}
\label{sec:adjGIFT}
In GIFT the windows necessary for sliding-window approach are created by defining a rectangular function over the whole time course of one session. Afterwards this vector is convolved by a Gaussian function. The resulting tapered window is then used to create windowed snapshots of the activity of RSNs used to build correlation matrices. In GIFT this procedure is implemented by correlating the whole windowed time courses including zero or near zero values, i.\,e. with sessions of length $T$ windowed time courses of length $T$ are correlated instead of restricting correlation to values above zero or above a small threshold. This procedure is valid, if the activity time courses within the windows have zero mean. But as soon as the deviation of the mean of activity time courses within a window is significantly large, the large number of zeros introduces exaggerated stability to the correlation coefficients.

Thus in standard dFNC analyses where the window length is sufficiently large to inhabit enough periods of the time courses to result in a zero mean distribution, the implemented approach is valid. But when dealing with time courses with averages deviating from zero within windows -- like it is the case when applying this approach to infra-slow frequency bands -- then GIFT implementation of dFNC has to be changed. We recommend to use pure box-car functions discarding the additional zeros just correlating the activity values of the windows per se or to use the tapered windows of the convolution and cut of values above a small threshold. Our extensive testing showed almost no difference between the two modifications. And as expected with decreasing frequency the deviation of the mean of the windowed time courses from zero increased. Thus the standard implementation still held for lower frequency bands, but above IMF index $f_8$ deviation was too large to guarantee valid results.

\renewcommand{\thefigure}{S\arabic{figure}}
\newpage

\subsection{Spectrum characteristics of ICs from gICA}
\setcounter{figure}{0}
\begin{figure}[htbp]
\centering
\includegraphics[width = .8\textwidth]{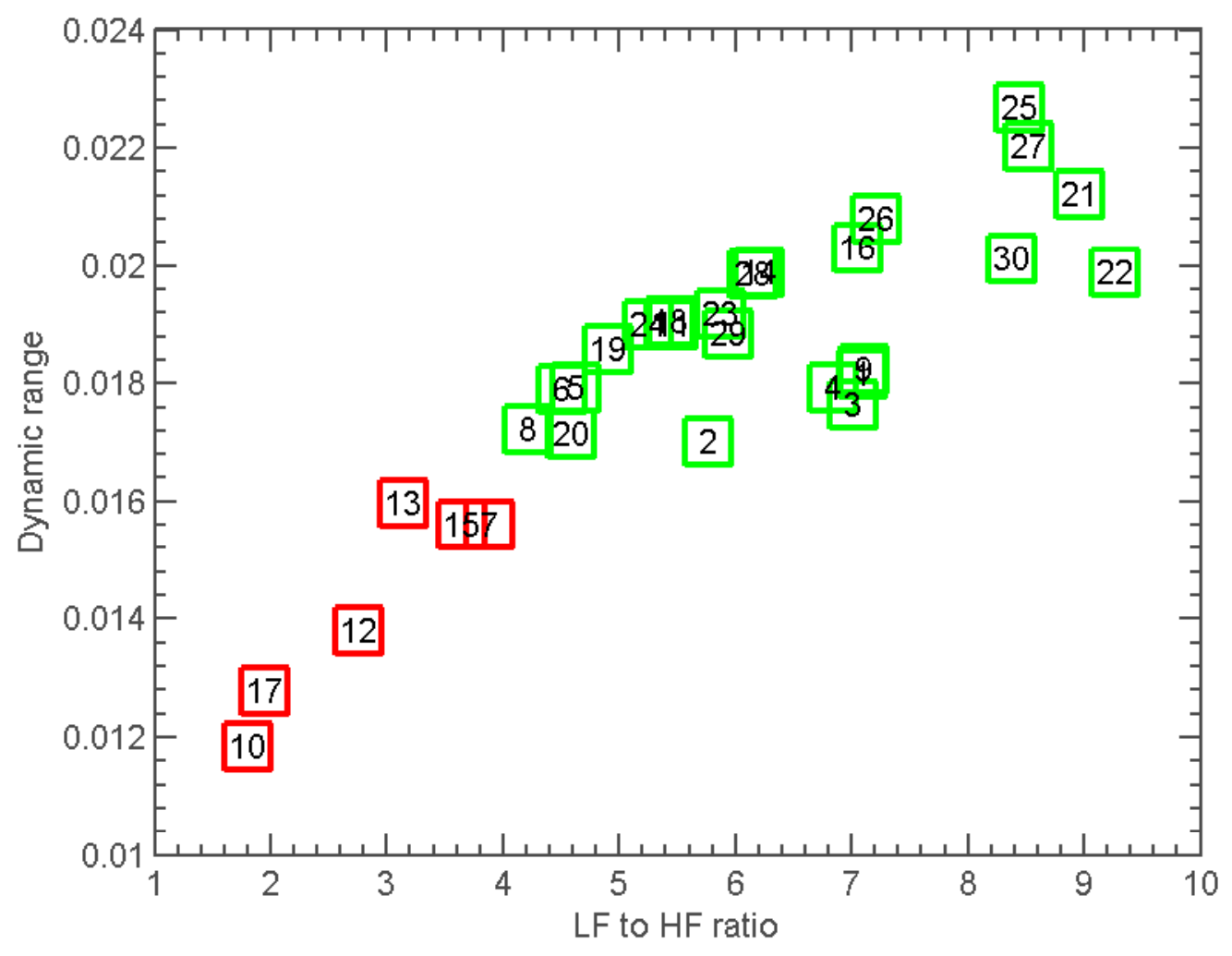}
\caption{In this figure the low frequency to high frequency ratio and the dynamic range are plotted for each IC resulting from gICA (numbers from fig. \ref{fig:gicaResEMD_appl}A are depicted). ICs with red boxes are discarded in the conservative data set. ICs 7, 15, and 17 are artifact ICs and it can be seen that ICs 10, 12, and 13 have worse spectrum characteristics than the best artifact IC 7. Therefore using the conservative data set is most valid.}
\label{fig:spectrumCharacteristics}
\end{figure}
\newpage
\subsection{Correlation and covariance between IMFs of different indices}
\begin{figure}[htbp]
\centering
\includegraphics{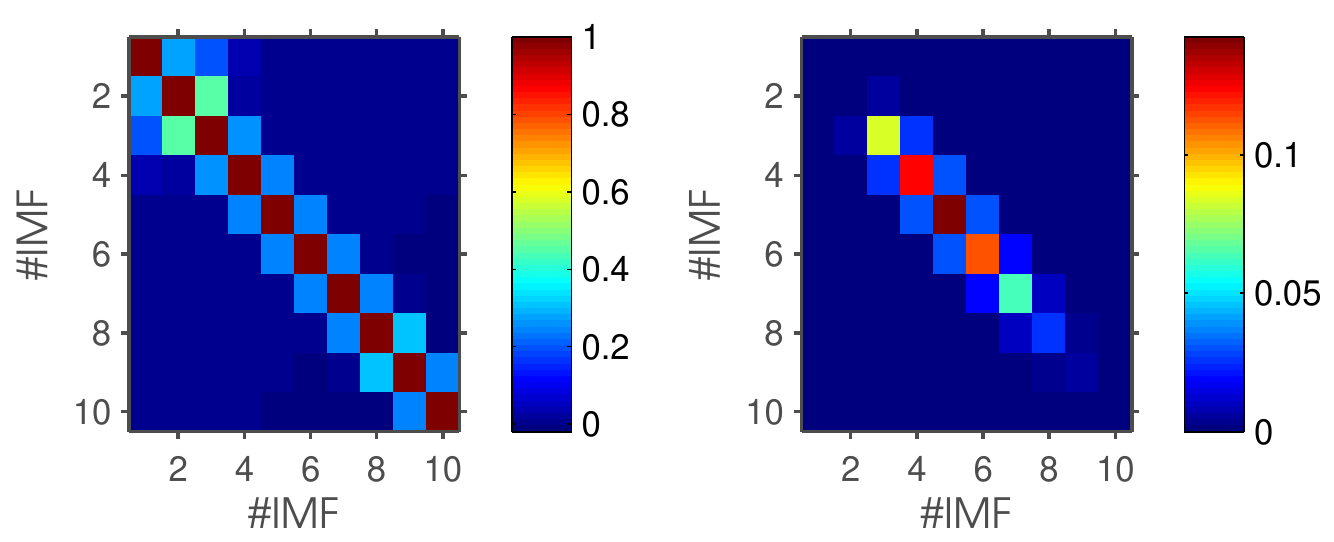}
\caption{Depicted are the average correlation matrix (left) and covariance matrix (right) of IMFs over sessions and components. The low correlation and covariance values confirm the separation in very narrow frequency bands.}
\label{fig:corrMatIMFs}
\end{figure}
\clearpage

\subsection{Scale-stability of connectivity-states for different $k$-means runs: original data}
\begin{figure}[htbp]
\centering
\includegraphics{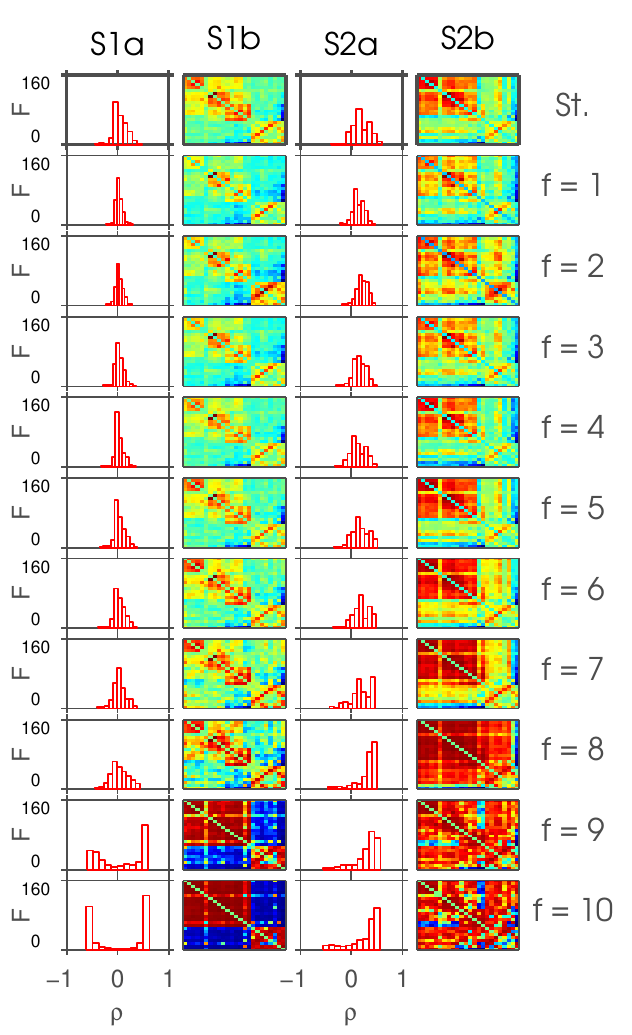}
\caption{Depicted is the result of the ordering algorithm \ref{alg:orderUnique} on the $k$-means run with $k = 2$. Connectivity-states are shown in columns with suffix *b and the coloring is individually adjusted to range from minimum to maximum value  to emphasize the structural similarity of connectivity states over frequency scales. The information of the distribution of correlation coefficients can be found in histograms plotted for corresponding connectivity-states in columns with suffix *a.}
\label{fig:scaleStability_robustnessK2}
\end{figure}
\begin{figure}[htbp]
\centering
\includegraphics{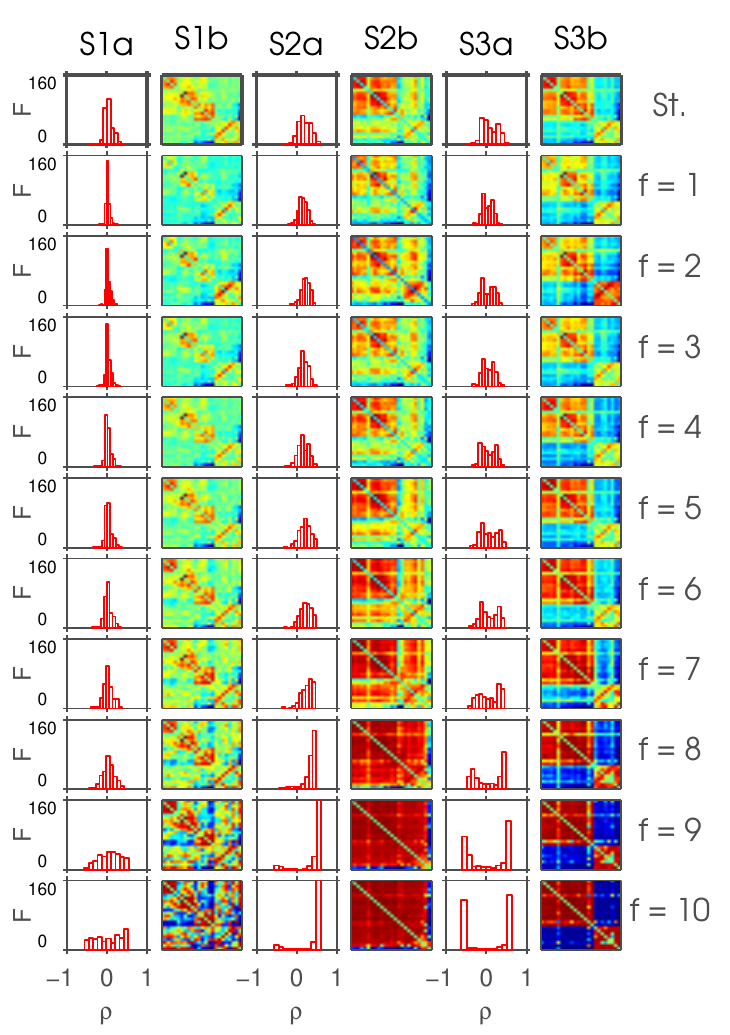}
\caption{Depicted is the result of the ordering algorithm \ref{alg:orderUnique} on the $k$-means run with $k = 3$. Connectivity-states are shown in columns with suffix *b and the coloring is individually adjusted to range from minimum to maximum value  to emphasize the structural similarity of connectivity states over frequency scales. The information of the distribution of correlation coefficients can be found in histograms plotted for corresponding connectivity-states in columns with suffix *a.}
\label{fig:scaleStability_robustnessK3}
\end{figure}
\begin{figure}[htbp]
\centering
\includegraphics[width = 1\textwidth]{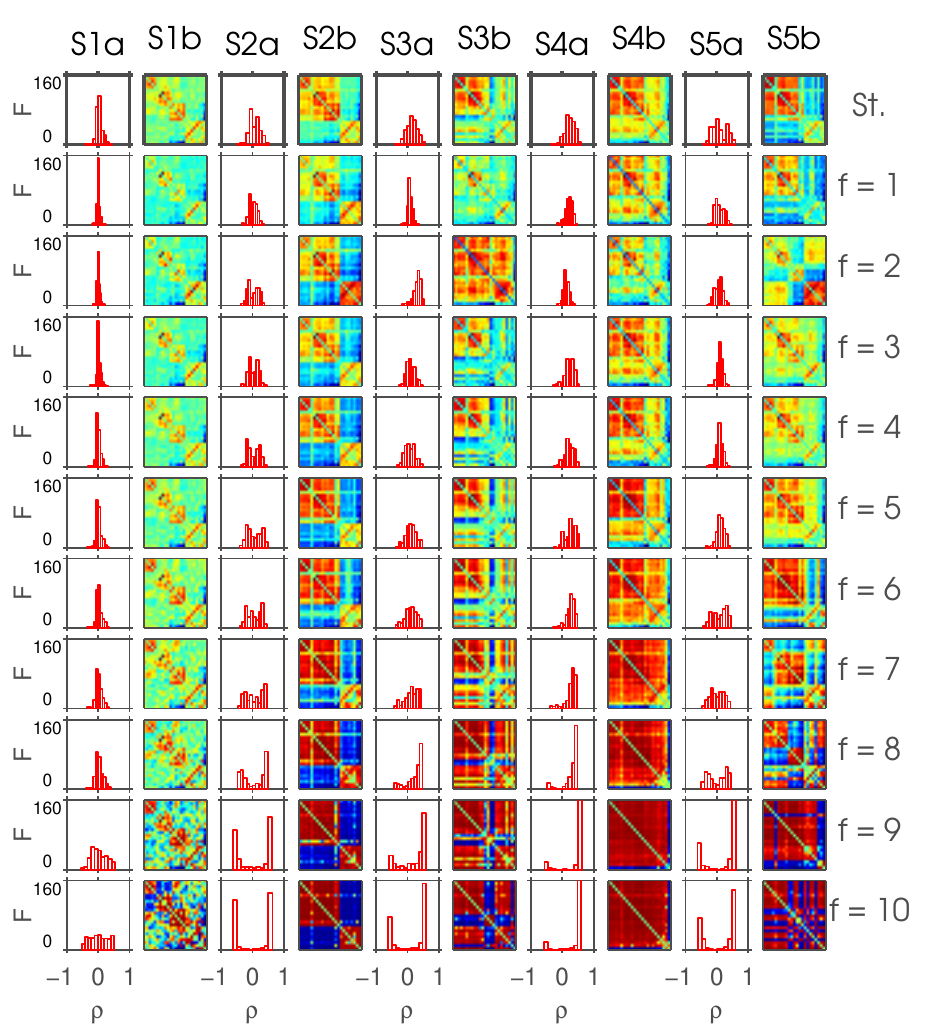}
\caption{Depicted is the result of the ordering algorithm \ref{alg:orderUnique} on the $k$-means run with $k = 5$. Connectivity-states are shown in columns with suffix *b and the coloring is individually adjusted to range from minimum to maximum value  to emphasize the structural similarity of connectivity states over frequency scales. The information of the distribution of correlation coefficients can be found in histograms plotted for corresponding connectivity-states in columns with suffix *a.}
\label{fig:scaleStability_robustnessK5}
\end{figure}

\begin{figure}[htbp]
\centering
\includegraphics[width = 1\textwidth]{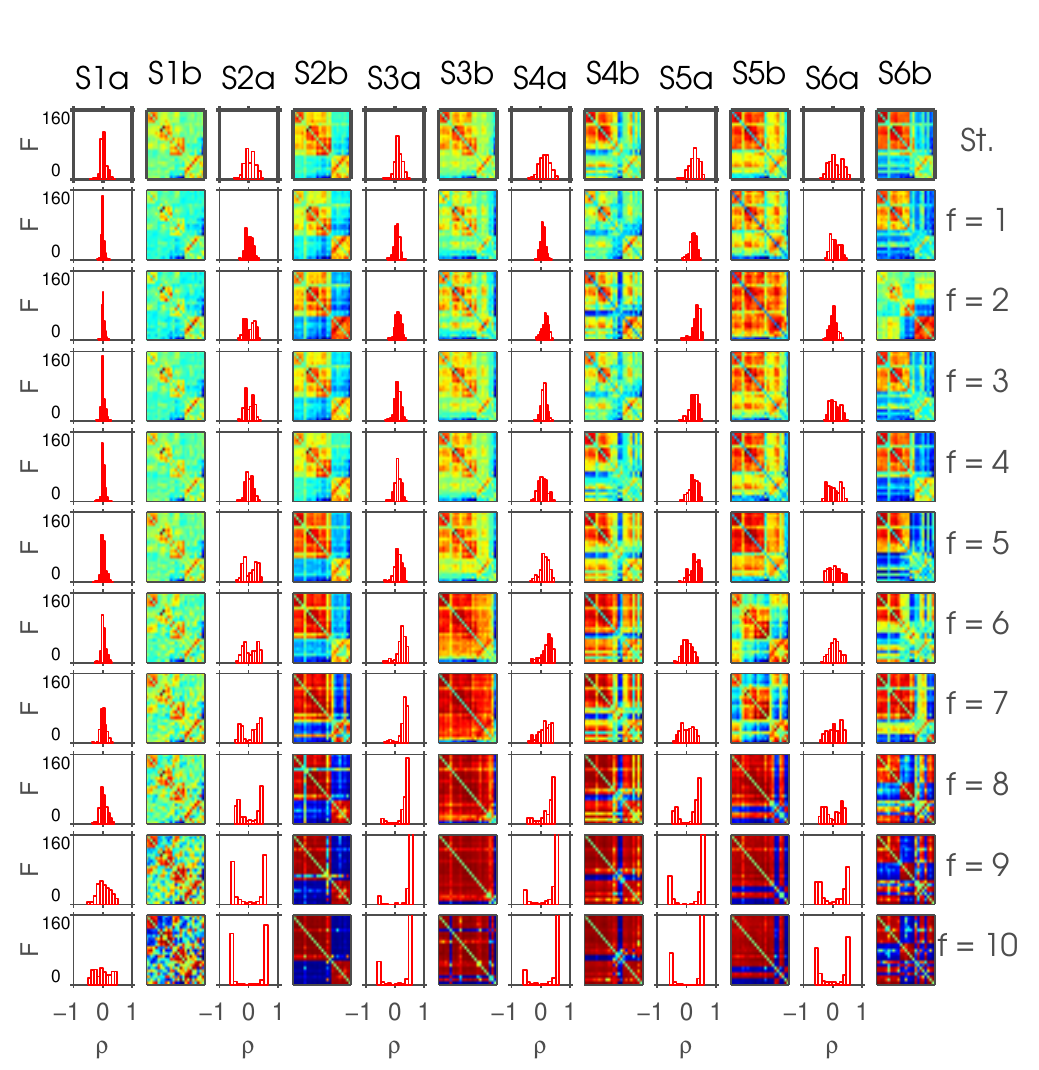}
\caption{Depicted is the result of the ordering algorithm \ref{alg:orderUnique} on the $k$-means run with $k = 6$. Connectivity-states are shown in columns with suffix *b and the coloring is individually adjusted to range from minimum to maximum value  to emphasize the structural similarity of connectivity states over frequency scales. The information of the distribution of correlation coefficients can be found in histograms plotted for corresponding connectivity-states in columns with suffix *a.}
\label{fig:scaleStability_robustnessK6}
\end{figure}

\begin{figure}[htbp]
\centering
\includegraphics[width = 1\textwidth]{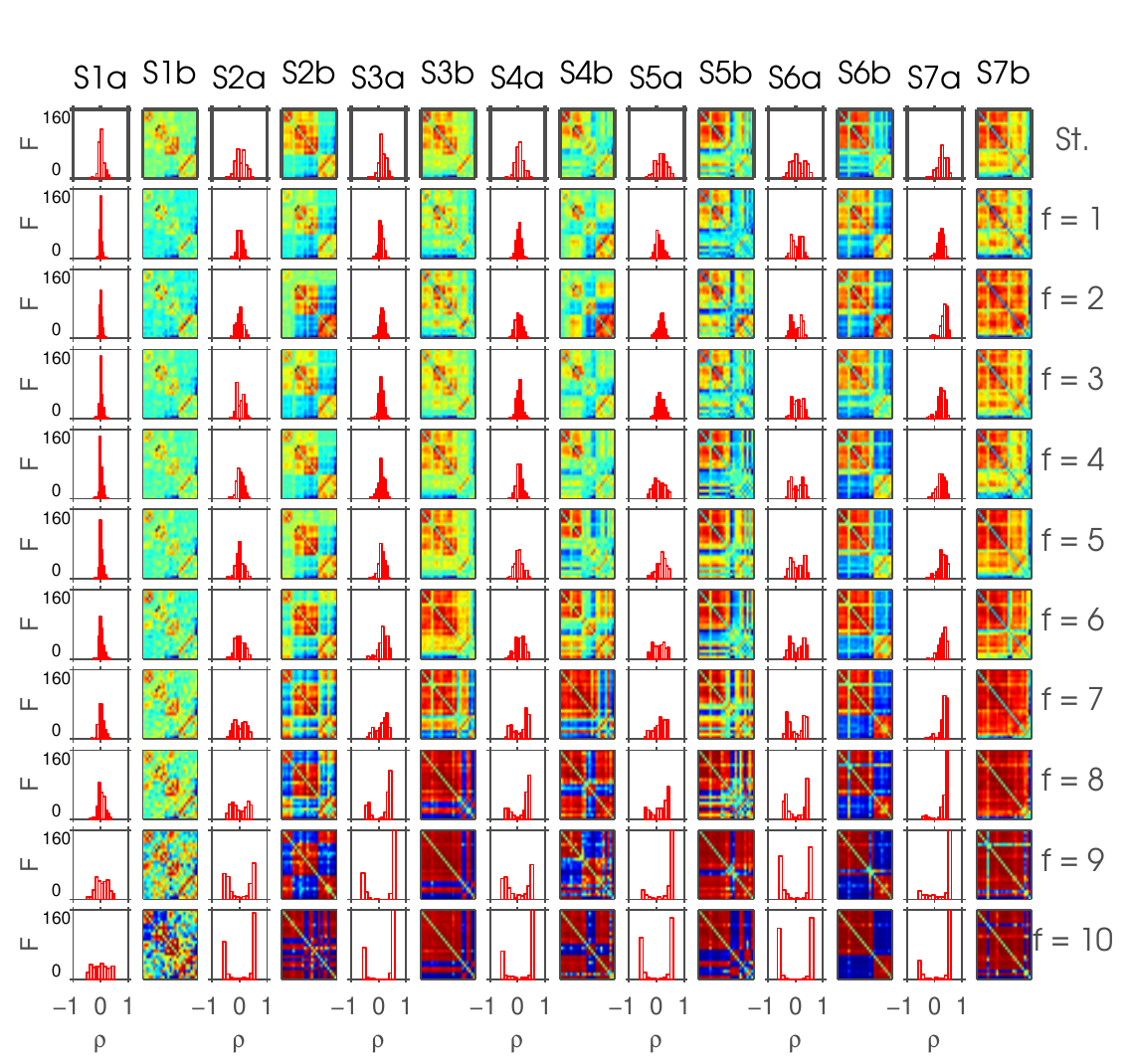}
\caption{Depicted is the result of the ordering algorithm \ref{alg:orderUnique} on the $k$-means run with $k = 7$. Connectivity-states are shown in columns with suffix *b and the coloring is individually adjusted to range from minimum to maximum value  to emphasize the structural similarity of connectivity states over frequency scales. The information of the distribution of correlation coefficients can be found in histograms plotted for corresponding connectivity-states in columns with suffix *a.}
\label{fig:scaleStability_robustnessK7}
\end{figure}
\begin{figure}[htbp]
\centering
\includegraphics[width = 1\textwidth]{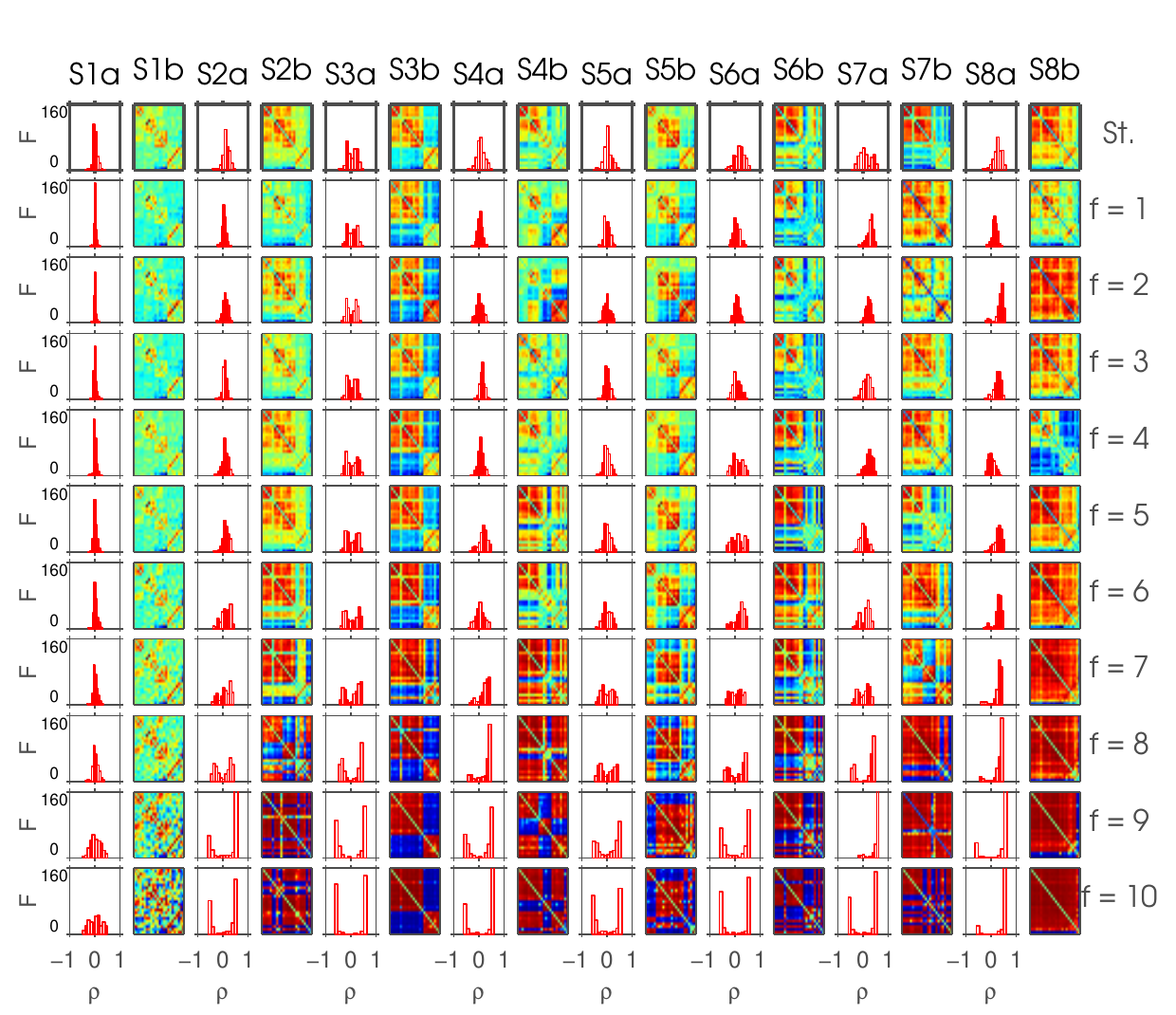}
\caption{Depicted is the result of the ordering algorithm \ref{alg:orderUnique} on the $k$-means run with $k = 8$. Connectivity-states are shown in columns with suffix *b and the coloring is individually adjusted to range from minimum to maximum value  to emphasize the structural similarity of connectivity states over frequency scales. The information of the distribution of correlation coefficients can be found in histograms plotted for corresponding connectivity-states in columns with suffix *a.}
\label{fig:scaleStability_robustnessK8}
\end{figure}
\begin{figure}[htbp]
\centering
\includegraphics[width = 1\textwidth]{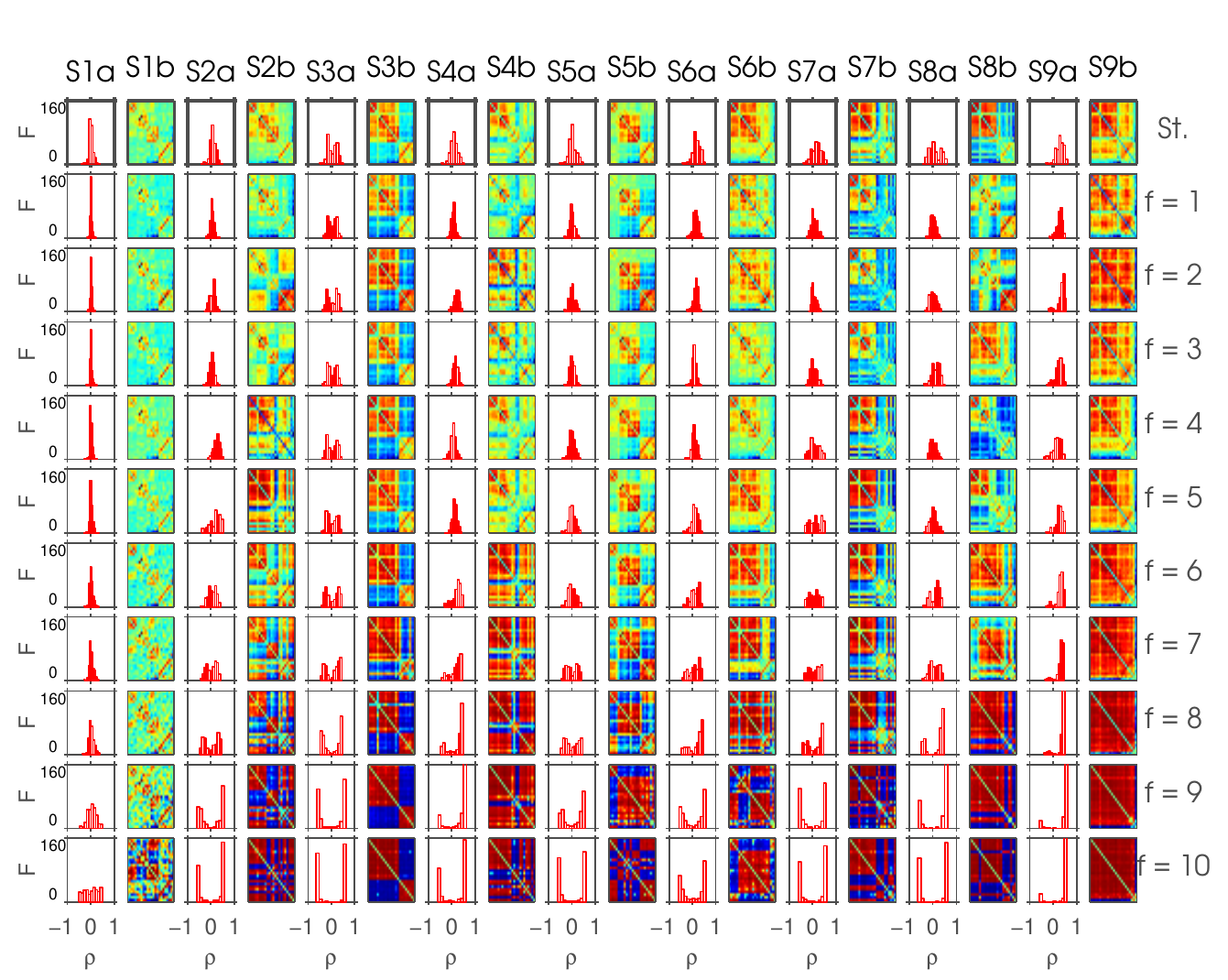}
\caption{Depicted is the result of the ordering algorithm \ref{alg:orderUnique} on the $k$-means run with $k = 9$. Connectivity-states are shown in columns with suffix *b and the coloring is individually adjusted to range from minimum to maximum value  to emphasize the structural similarity of connectivity states over frequency scales. The information of the distribution of correlation coefficients can be found in histograms plotted for corresponding connectivity-states in columns with suffix *a.}
\label{fig:scaleStability_robustnessK9}
\end{figure}

\begin{figure}[htbp]
\centering
\includegraphics[width = 1\textwidth]{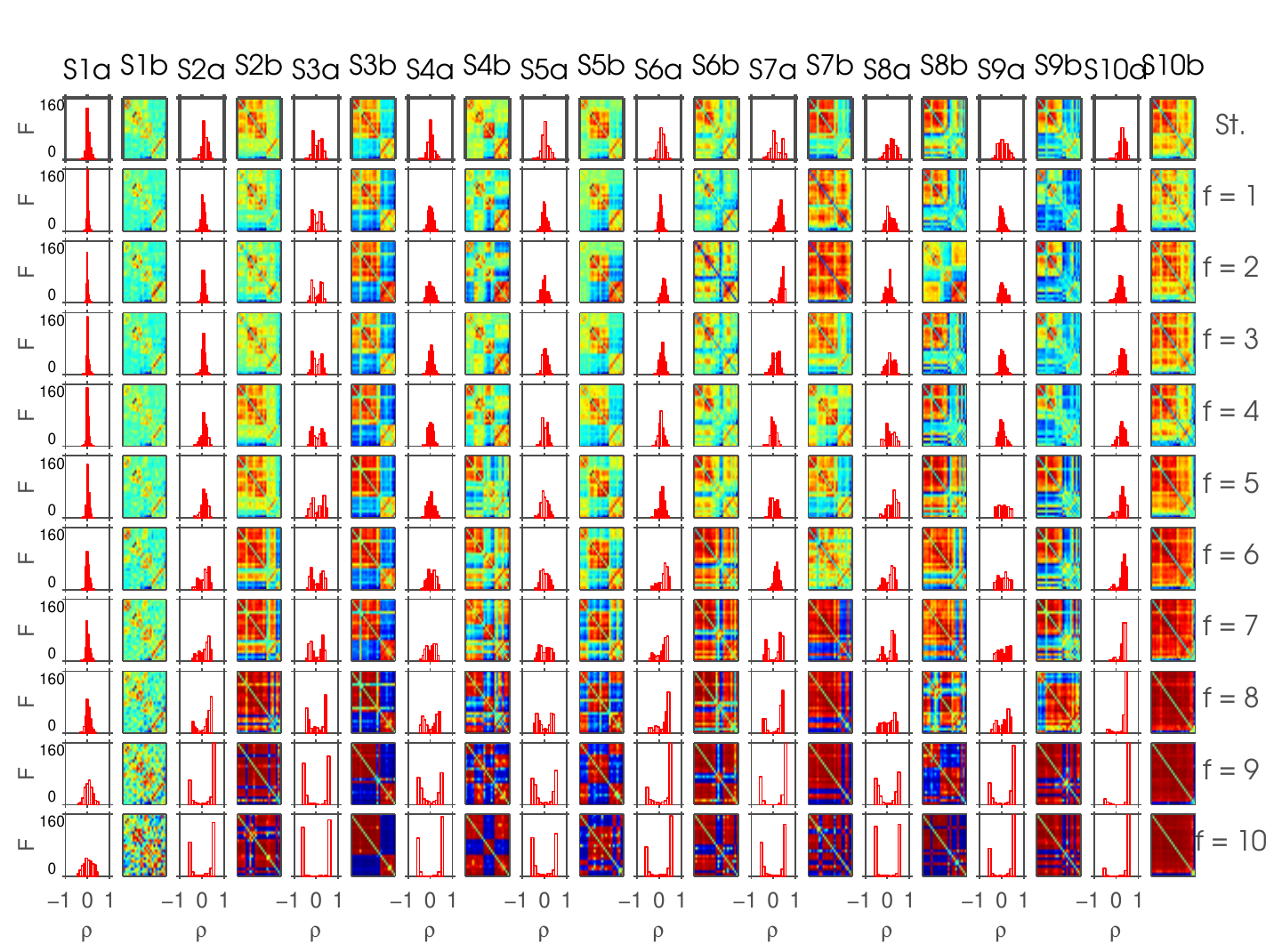}
\caption{Depicted is the result of the ordering algorithm \ref{alg:orderUnique} on the $k$-means run with $k = 10$. Connectivity-states are shown in columns with suffix *b and the coloring is individually adjusted to range from minimum to maximum value  to emphasize the structural similarity of connectivity states over frequency scales. The information of the distribution of correlation coefficients can be found in histograms plotted for corresponding connectivity-states in columns with suffix *a.}
\label{fig:scaleStability_robustnessK10}
\end{figure}
\clearpage

\subsection{One exemplar of $k$-means applied on component-wise temporally demeaned dFC matrices}
\begin{figure}[h!]
\centering
\includegraphics[width = .7\textwidth]{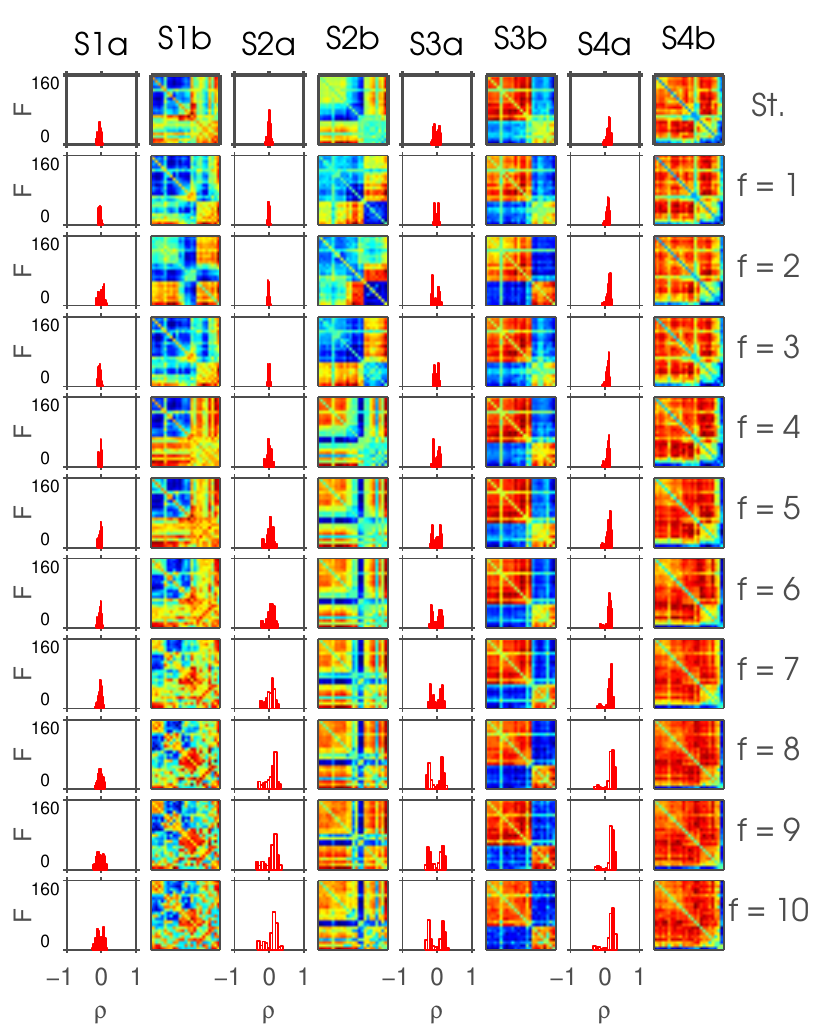}
\caption{Depicted is the result of the ordering algorithm \ref{alg:orderUnique} on the $k$-means run with $k = 4$ applied on the demeaned version of the canonical approach. Connectivity-states are shown in columns with suffix *b and the coloring is individually adjusted to range from minimum to maximum value. The information of the distribution of correlation coefficients can be found in histograms plotted for corresponding connectivity-states in columns with suffix *a. Here eleven frequency scales are shown, since the MEMD algorithm extracted more IMFs than in the original time courses. This is due to the fact that shuffling introduces higher frequencies to the time courses.}
\label{fig:scaleStability_robustnessK4_demeaned}
\end{figure}
\clearpage

\subsection{One exemplar of $k$-means applied on shuffled time courses over frequency scales}
\begin{figure}[h!]
\centering
\includegraphics[width = .7\textwidth]{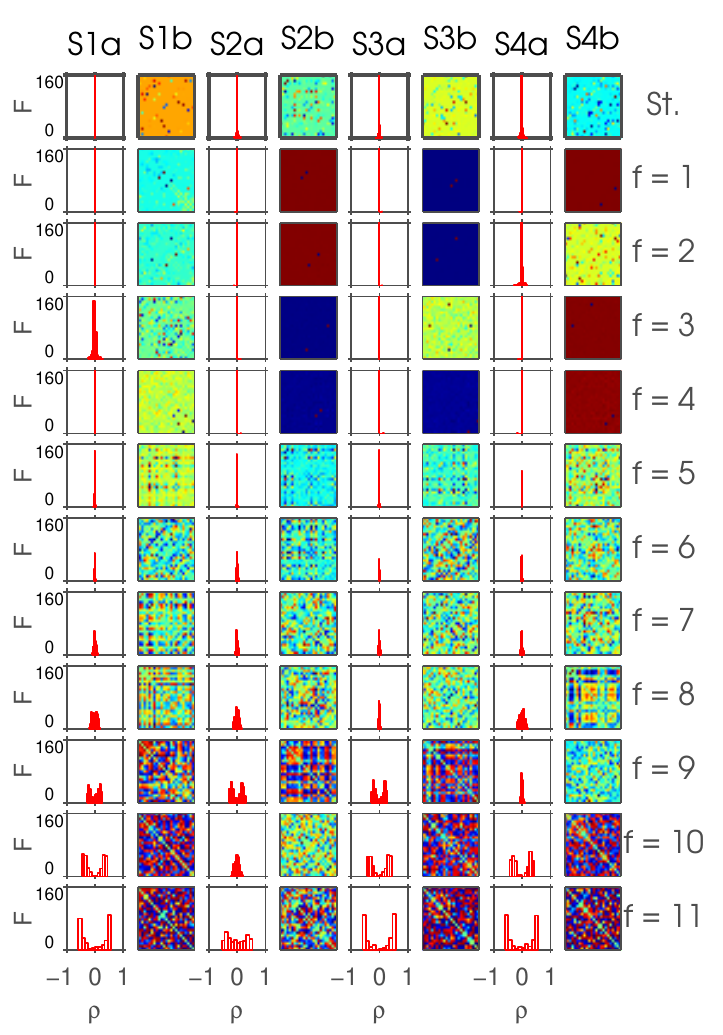}
\caption{Depicted is the result of the ordering algorithm \ref{alg:orderUnique} on the $k$-means run with $k = 4$ applied on the shuffled version of the original time courses. Connectivity-states are shown in columns with suffix *b and the coloring is individually adjusted to range from minimum to maximum value. The information of the distribution of correlation coefficients can be found in histograms plotted for corresponding connectivity-states in columns with suffix *a. Here eleven frequency scales are shown, since the MEMD algorithm extracted more IMFs than in the original time courses. This is due to the fact that shuffling introduces higher frequencies to the time courses.}
\label{fig:scaleStability_robustnessK4_shuffled}
\end{figure}
\clearpage

\subsection{One exemplar of $k$-means applied on phase-randomized time courses over frequency scales}
\begin{figure}[h!]
\centering
\includegraphics[width = .7\textwidth]{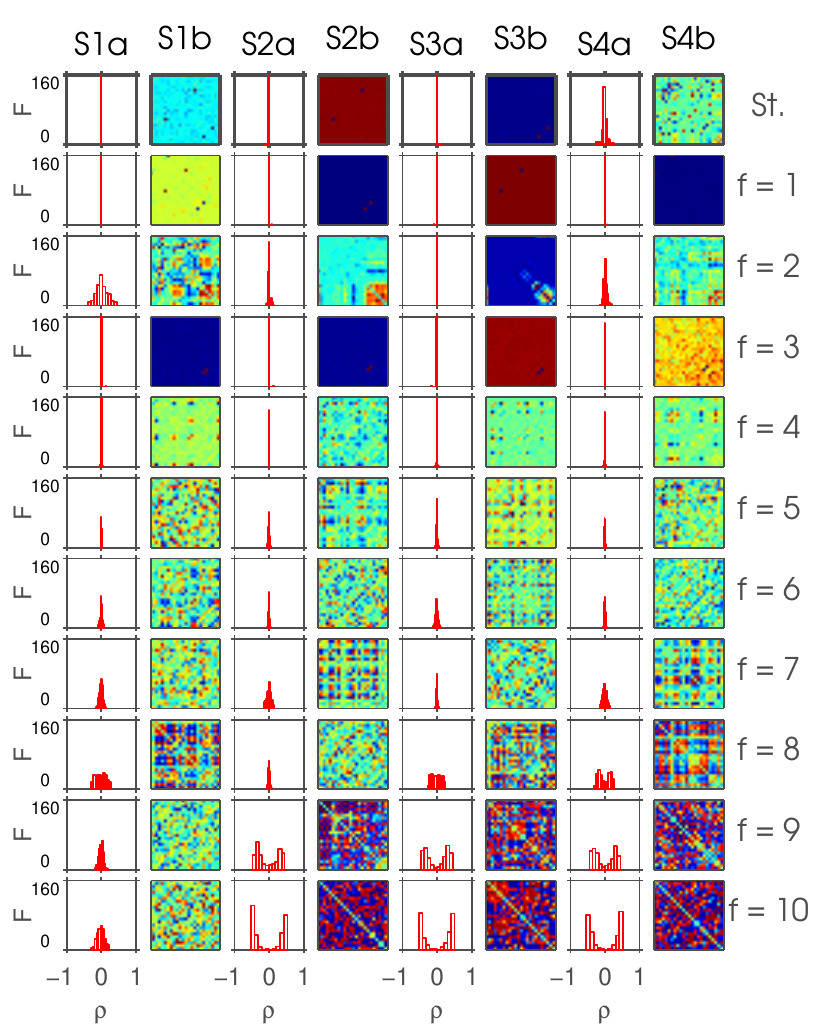}
\caption{Depicted is the result of the ordering algorithm \ref{alg:orderUnique} on the $k$-means run with $k = 4$ applied on the phase-randomized version of the original time courses. Connectivity-states are shown in columns with suffix *b and the coloring is individually adjusted to range from minimum to maximum value. The information of the distribution of correlation coefficients can be found in histograms plotted for corresponding connectivity-states in columns with suffix *a.}
\label{fig:scaleStability_robustnessK4_phaseRand}
\end{figure}
\clearpage

\subsection{Complexity analysis of real and simulated data}
\label{sec:iMMSE_sim}

\begin{figure}[h!tbp]
\centering
\includegraphics[width = 1\textwidth]{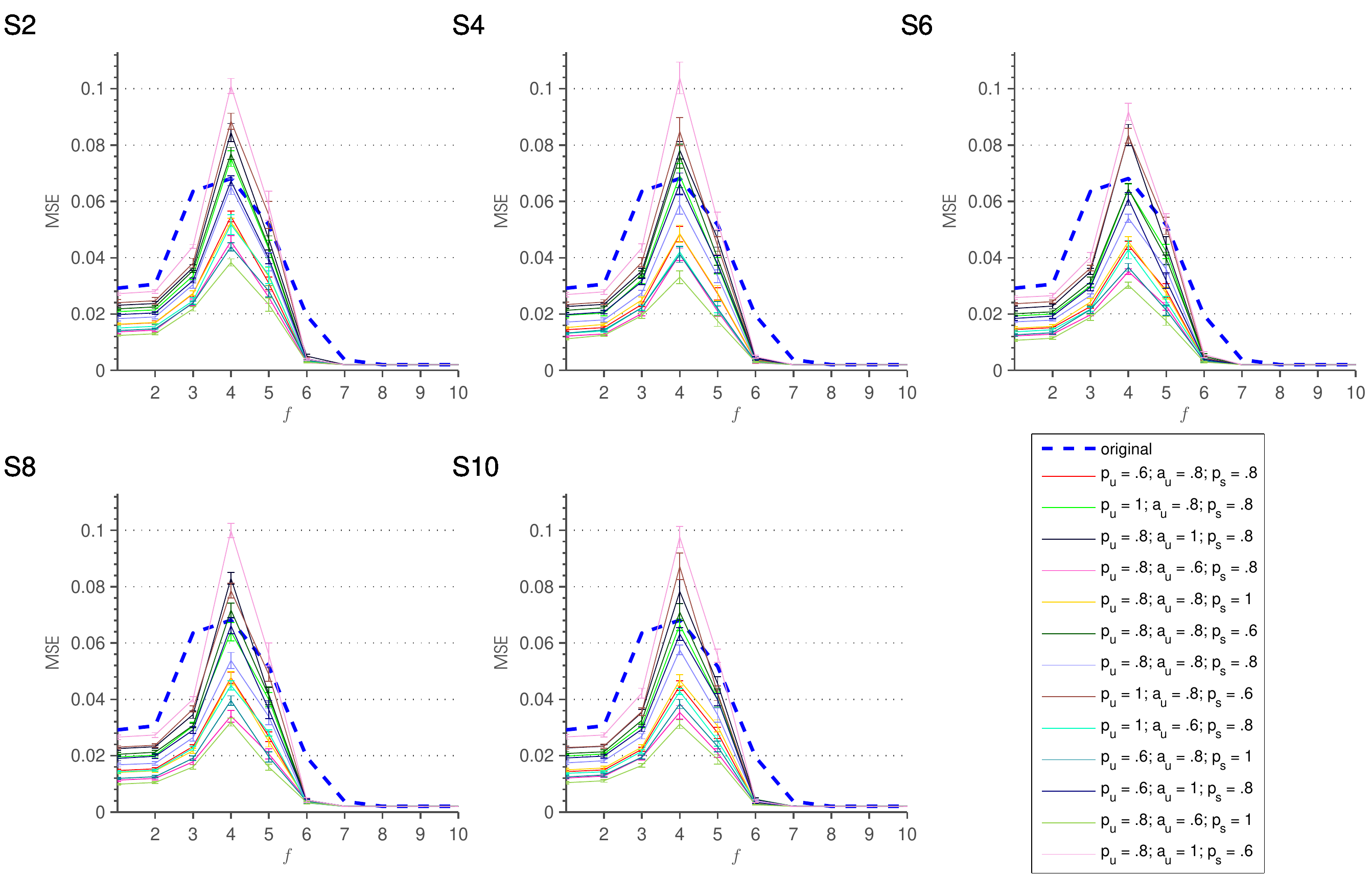}
\caption{These plots show the results of the complexity analysis of the real and simulated data in terms of iMMSE. Panels S2-10 correspond to the number of simulated connectivity-states $k_{sim}$. The blue dashed line shows iMMSE of averaged over all $400$ sessions. For each simulation run iMMSE was calculated for the first ten 'sessions'. Errorbars represent the standard error. The legend shows the parameters used for a certain simulation.}
\label{fig:iMMSE_sim}
\end{figure}
\clearpage

\subsection{Adjustable window size}
\begin{figure}[h!tbp]
\centering
\includegraphics[width = .5\textwidth]{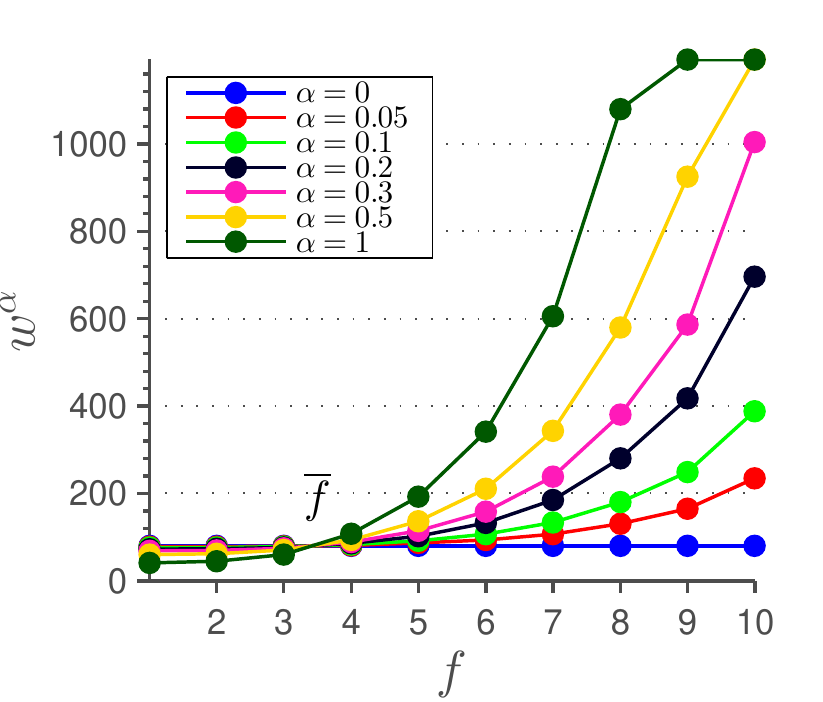}
\caption{This figure shows the adjustment of window size $w^{\alpha}$ on frequency scales $f$ used for the sliding-window approach. The $\alpha$-values range from the one extreme case with constant window size ($\alpha = 0$) to the case with a constant average number of periods in each window ($\alpha = 1$).}
\label{fig:adjW}
\end{figure}
\clearpage

\subsection{Filter-banks (constant order): rs-fMRI data}
\begin{figure}[h!tbp]
\centering
\includegraphics[width = .8\textwidth]{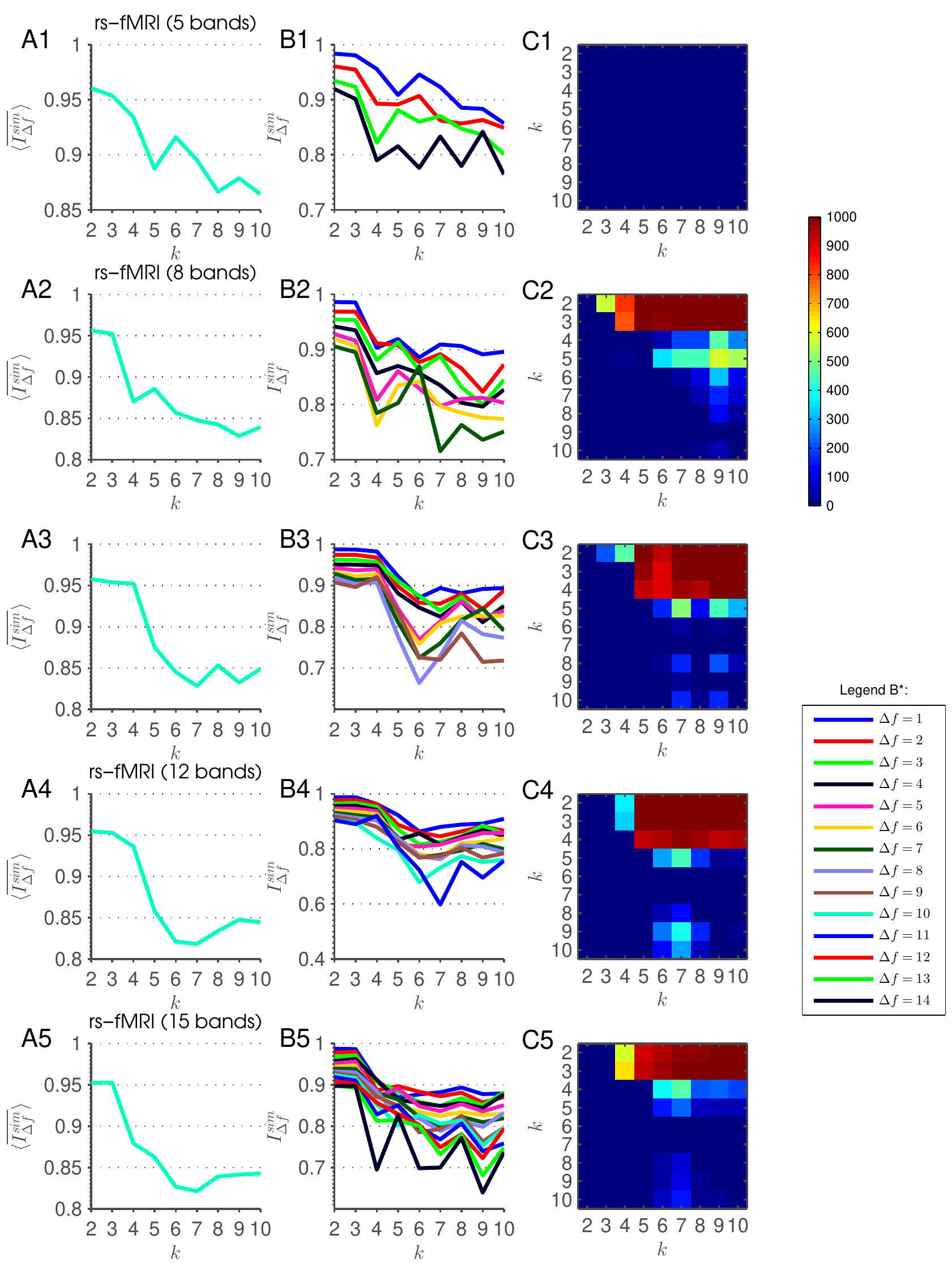}
\caption{This figure summarizes the results of the scale-stability analysis for the filter-bank procedures with constant filter order applied on rs-fMRI data: (*1) 5 bands; (*2) 8 bands; (*3) 10 bands; (*4) 12 bands; (*5) 15 bands. The significance levels are adjusted to $\alpha = .05$ (*1) and (*3), $\alpha = .00007$ (*2) and (*5), $\alpha = .00196$ (*4), and $\alpha = .00008$ (*6).}
\label{fig:scaleStabilityIndex_filterBanks_realTCs_new}
\end{figure}
\clearpage

\subsection{Filter-Banks (constant order): simulated data}
\begin{figure}[h!tbp]
\centering
\includegraphics[width = .8\textwidth]{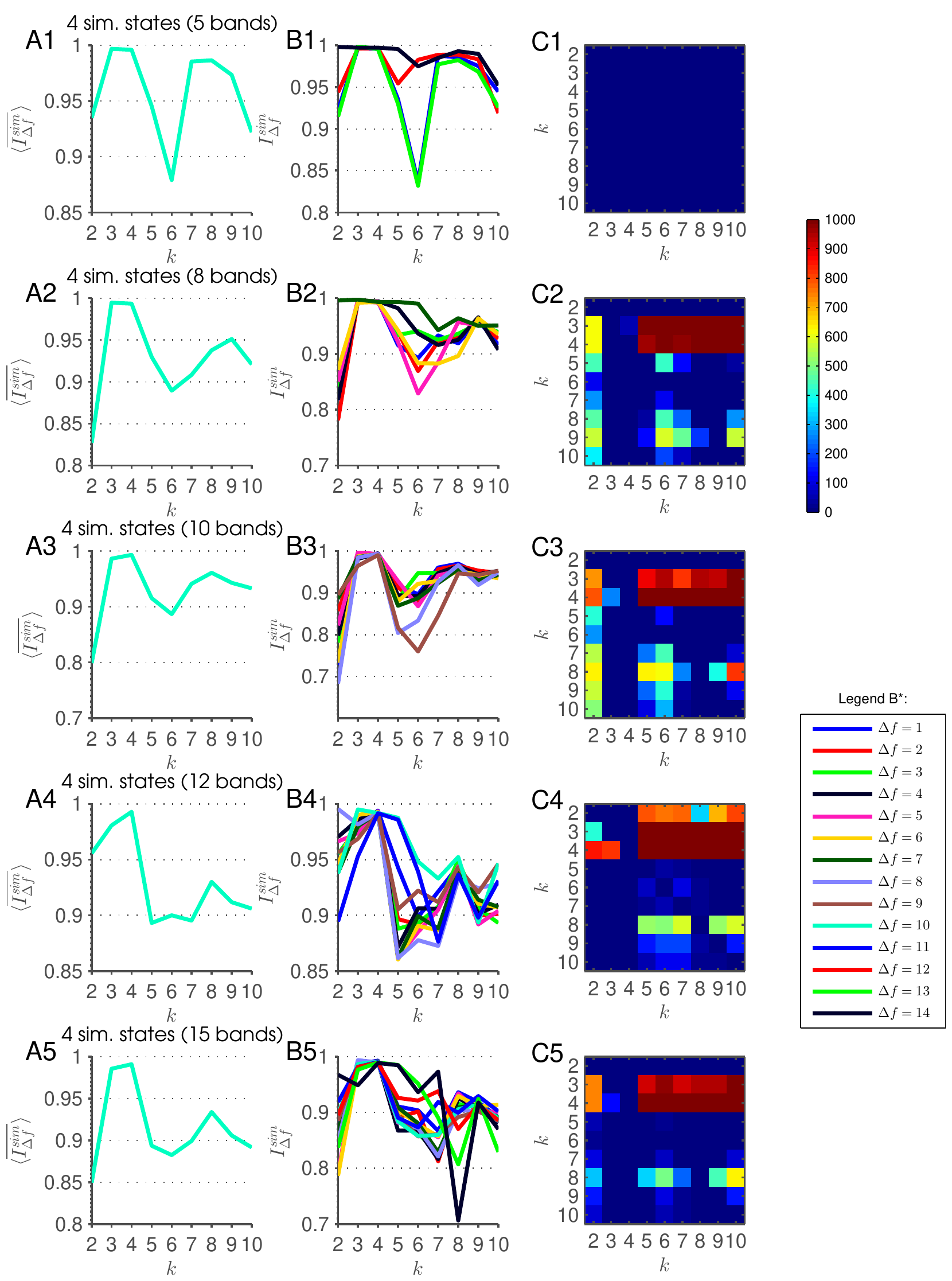}
\caption{This figure summarizes the results of the scale-stability analysis for the filter-bank procedures applied on rs-fMRI data: (*1) 5 bands; (*2) 8 bands; (*3) 10 bands; (*4) 12 bands; (*5) 15 bands. The significance levels are adjusted to $\alpha = .05$ (*1) and (*3), $\alpha = .00007$ (*2) and (*5), $\alpha = .00196$ (*4), and $\alpha = .00008$ (*6).}
\label{fig:scaleStabilityIndex_filterBanks_simTCs_new}
\end{figure}
\clearpage

\subsection{Filter-banks (adjusted order): rs-fMRI data}
\begin{figure}[h!tbp]
\centering
\includegraphics[width = .8\textwidth]{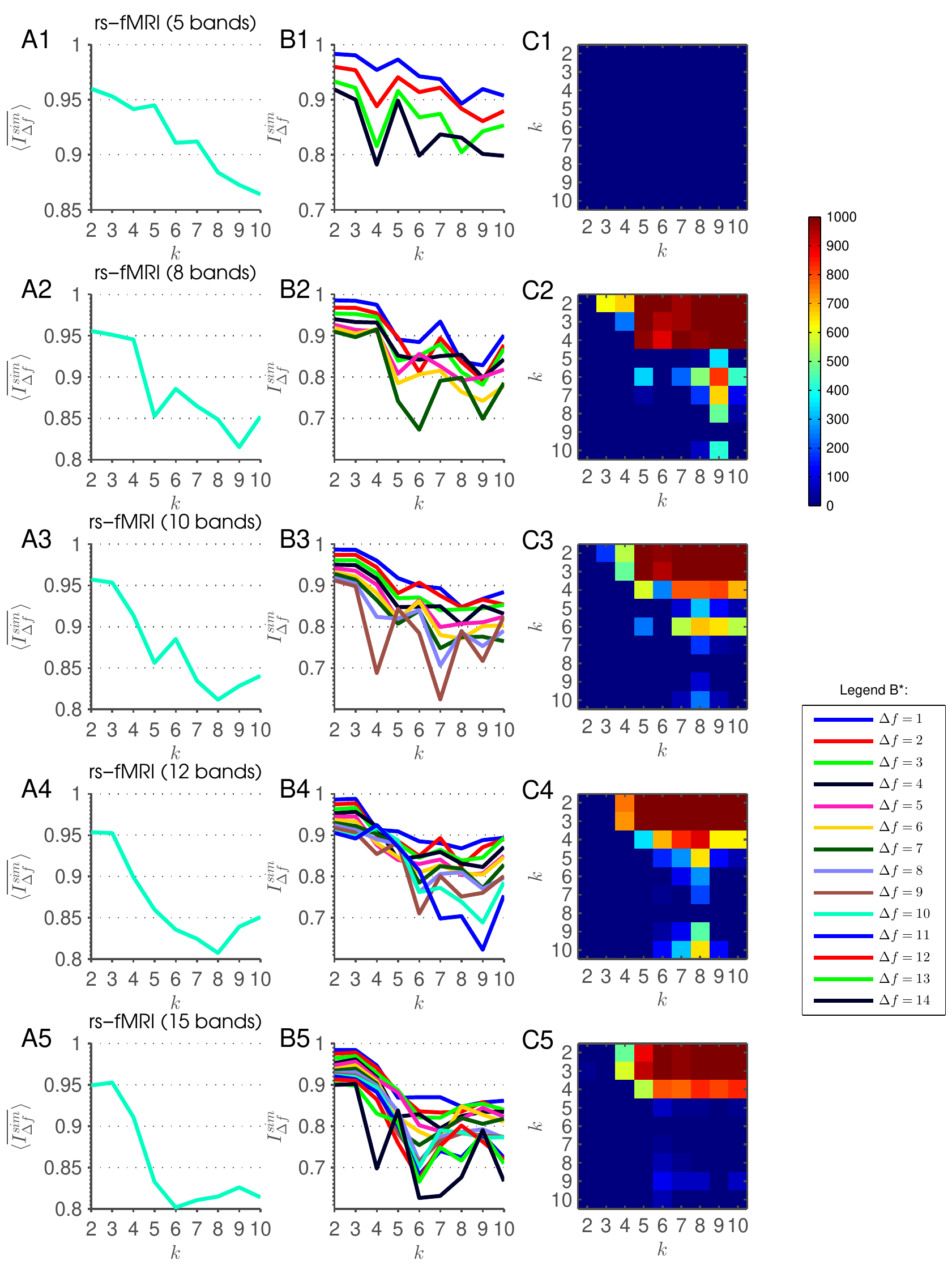}
\caption{This figure summarizes the results of the scale-stability analysis for the filter-bank procedures applied on rs-fMRI data: (*1) 5 bands; (*2) 8 bands; (*3) 10 bands; (*4) 12 bands; (*5) 15 bands. The significance levels are adjusted to $\alpha = .05$ (*1) and (*3), $\alpha = .00007$ (*2) and (*5), $\alpha = .00196$ (*4), and $\alpha = .00008$ (*6).}
\label{fig:scaleStabilityIndex_filterBanks_realTCs_new2}
\end{figure}
\clearpage

\subsection{Filter-Banks (adjusted order): simulated data}
\begin{figure}[h!tbp]
\centering
\includegraphics[width = .8\textwidth]{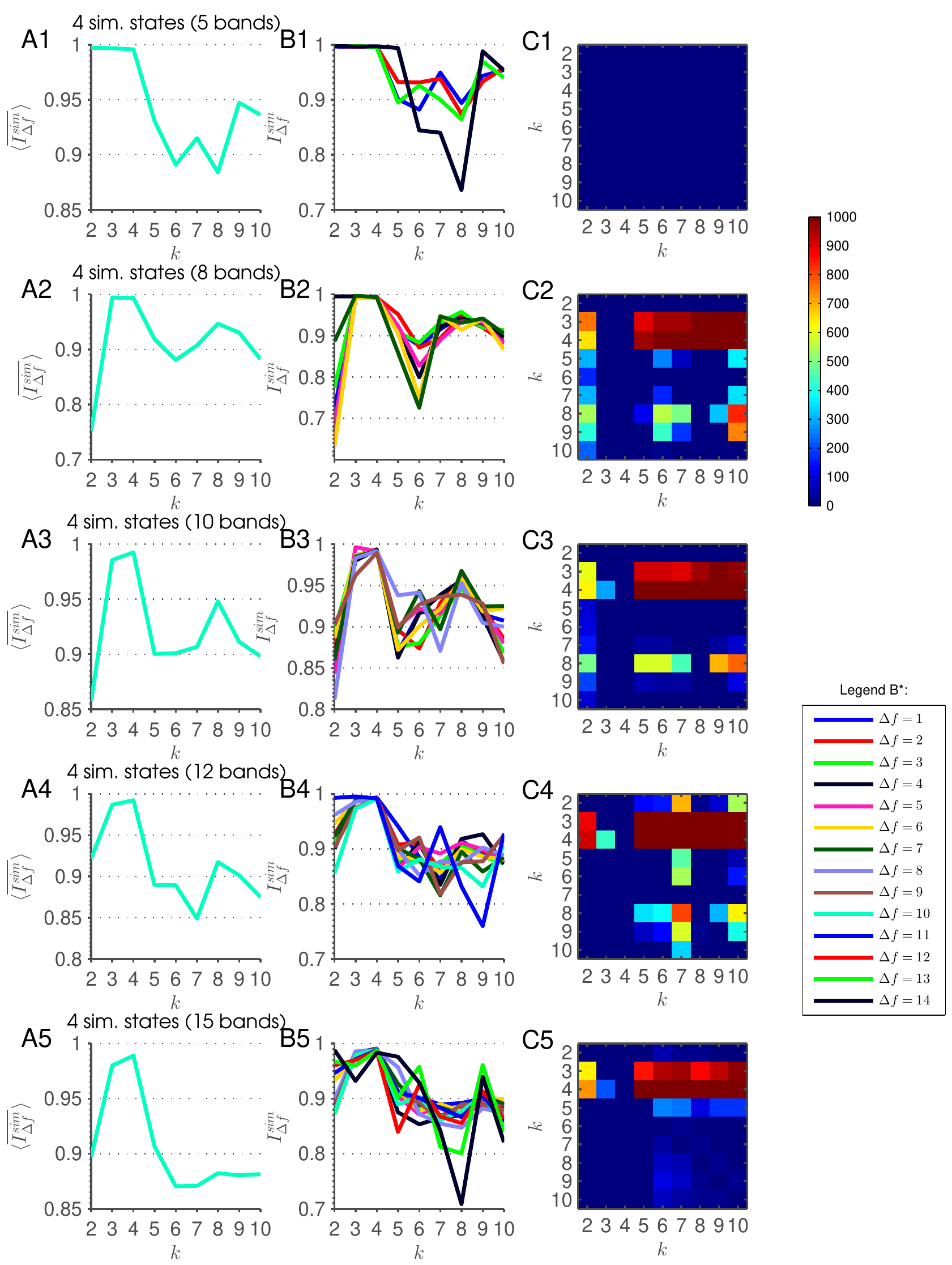}
\caption{This figure summarizes the results of the scale-stability analysis for the filter-bank procedures applied on rs-fMRI data: (*1) 5 bands; (*2) 8 bands; (*3) 10 bands; (*4) 12 bands; (*5) 15 bands. The significance levels are adjusted to $\alpha = .05$ (*1) and (*3), $\alpha = .00007$ (*2) and (*5), $\alpha = .00196$ (*4), and $\alpha = .00008$ (*6).}
\label{fig:scaleStabilityIndex_filterBanks_simTCs_new2}
\end{figure}
\clearpage

\end{document}